\documentclass[12pt,onecolumn, draftcls]{IEEEtran}
\usepackage[cmex10]{amsmath}
\usepackage{cite}
\usepackage{algorithm}
\usepackage{algorithmic}
\usepackage{array}
\usepackage{mdwtab}
\usepackage[tight,footnotesize]{subfigure}
\usepackage[dvips]{graphicx}
\usepackage{subfigure}
\usepackage{amsfonts}
\usepackage{makecell}
\newcommand\mgape[1]{\gape{$\vcenter{\hbox{#1}}$}}

\begin{document}

\graphicspath{{eps/}}

\DeclareGraphicsExtensions{.eps}
\title{Fast Compressed Sensing SAR Imaging based on Approximated Observation}
%
%
%

\author{Jian Fang, Zongben Xu, Bingchen Zhang, Wen Hong, Yirong Wu
\thanks{J. Fang and Z. B. Xu are with the School of Mathematics and Statistics, Xi'an Jiaotong University, Xi'an 710049, China}
\thanks{B. C. Zhang, W. Hong, Y. R. Wu are with the Institute of Electronics, Chinese Academy of Sciences, Beijing 100190, China}
}

\maketitle

\begin{abstract}
In recent years, compressed sensing (CS) has been applied in the field of synthetic aperture radar (SAR) imaging and shows great potential. The existing models are, however, based on application of the sensing matrix acquired by the exact observation functions. As a result, the corresponding reconstruction algorithms are much more time consuming than traditional matched filter (MF) based focusing methods, especially in high resolution and wide swath systems. In this paper, we formulate a new CS-SAR imaging model based on the use of the approximated SAR observation deducted from the inverse of focusing procedures. We incorporate CS and MF within an sparse regularization framework that is then solved by a fast iterative thresholding algorithm. The proposed model forms a new CS-SAR imaging method that can be applied to high-quality and high-resolution imaging under sub-Nyquist rate sampling, while saving the computational cost substantially both in time and memory. Simulations and real SAR data applications support that the proposed method can perform SAR imaging effectively and efficiently under Nyquist rate, especially for large scale applications.
\end{abstract}


%
\IEEEpeerreviewmaketitle

\section{Introduction}
%
%
%
%
\IEEEPARstart{S}{ynthetic} aperture radar (SAR) is an active microwave radar which can achieve high-resolution images in all time of day and  weather \cite{Cumming2004}. In a SAR system, the radar emits a sequence of pulses along its path and receives the echoes (raw data) scattered from the targets. The reconstruction of the scene is traditionally achieved by matched filter (MF) based focusing algorithms which are efficient but need Nyquist rate samples of the echoes. The SAR imaging with increasing resolution and swath requires more and more measurements, storage and downlink bandwidth. The current system hardware, however, frequently hampers such high-dimensional application.

The recent development of compressed sensing (CS) brings possibility of reconstructing sparse or compressible signals with fewer measurements than that Nyquist requires\cite{Candes2006cs,Baraniuk2007,Donoho2006}. Several applications on radar system appear in recent years, which primarily concern how the data acquisition way can be simplified by using CS\cite{Gurbuz2009}\cite{Herman2009} and what the potential applications will renovate radar imaging with CS technique\cite{Ender2010}\cite{Potter2010}. Further, in the study of CS-SAR, much attention has been paid to the effective use of the specific SAR geography and signal form, say, in \cite{Bhattacharya2007}, a SAR raw data compression framework based on CS was suggested by sampling the data in frequency domain. An extension of this work was given in \cite{Rilling2009} by using the fact that very bright objects are always sparse, resulting in a hybrid sparse model. These works, however, do not apply to the CS-SAR system practically where sampling is expected in time-domain. In \cite{Tello2010}, CS was applied on azimuth after the range compression. By combining range MF, the method was much more efficient, while, the redundant information in range has not been effectively utilized. More general CS-SAR model were reported in \cite{Patel2010}\cite{Zeng2011} by discretizing the SAR observation function exactly into an observation matrix, while solving by CS straightforwardly.

All those works strongly demonstrated that some exclusive advantages of CS-SAR do exist as compared to the traditional SAR imaging methodologies, say, relaxation of required measurements, reduction of side lobe and a further suppression of noise \cite{Cetin2001}. However, in all applications, a serious drawback has been observed: as compared to the traditional MF based methods, the computational complexity and memory cost of the CS-SAR models are much higher, so that it is very inefficient to be applied to high-dimensional applications.

In this paper, we formulate a new CS-SAR framework within which the computational complexity of the  CS-SAR imaging can be significantly reduced. Our main idea is to replace the exact observation function in the CS-SAR framework with approximated observations derived from the inverse of traditional MF based procedures. Such inversion has ever been applied to yield raw signals (the echoes) in a more economical way \cite{Khwaja2005}\cite{Franceschetti2004}, but requires high accuracy of the adopted method. In this paper, we take a further step by incorporating it into the CS framework, which demands only a well focusing ability to ensure CS reconstruction. We propose to implement the CS-SAR imaging through the sparse regularization scheme which is then solved by an iterative thresholding algorithm (ITA). Accordingly, the fast speed and high efficiency of the new method are guaranteed respectively from the use of the approximated observation and from thes CS reconstruction procedure. We show that the new CS-SAR imaging method can not only acquire high-quality and high-resolution images with significantly reduced measurements, but also reduces the memory cost to $\mathcal{O}(n)$ and computational complexity of one-step iteration to $\mathcal{O}(n\log n)$, achieving the same order with the traditional SAR imaging methods.

The reminder of the paper is organized as follows. In Section 2, we introduce the background knowledge on the stripmap mode SAR system and the classical CS-SAR model.  In section 3, we present the approximated observation by calculating the inverse of MF imaging procedure. In Section 4, we formulate the new CS-SAR imaging method through hybridizing the approximated observation and sparse regularization. In Section 5, we show the simulation and application results of the suggested method. Conclusions are then presented in Section 6 with some useful remarks.

\emph{Notation:} We will use the subsequent notations throughout the paper: Column vectors, matrices and operators will be denoted respectively by bold lower case, $\bf x$, bold upper case, $ \textbf{A}$, and roman upper case, $\rm C$. $ \textbf{A}^{\bf T},\textbf{A}^\ast , \textbf{A}^{\bf H}$ denotes the transpose, conjugate and Hermitian transpose of $\bf A$, respectively.




\section{CS-SAR models based on exact observation}
In this section, some preliminary knowledge of CS-SAR imaging is summarized. We focus on the general formalization of CS-SAR model, with a more detailed introduction of the iterative thresholding procedures for solution of the CS-SAR models.

\subsection{Stripmap Mode SAR Model}
In the stripmap mode SAR, the antenna is pointed to a fixed direction and the platform flights with constant velocity $v$. Then, a complex baseband $p_c(\tau)$, usually chirp, is modulated to real pulse $p(\tau)=\cos(2\pi f_0\tau+\phi(\tau))(-\frac{t_s}{2}\le\tau\le \frac{t_s}{2})$ ($f_0$ is the carrier frequency, $\tau$ is the range time, $W_r$ is the elevation weight and $t_s$ is the pulse duration) and transmitted at a constant pulse repetition frequency (PRF). The received backscattered energy can then be modeled as a convolution of the pulse waveform with the ground reflectivity function, given by \cite{Wu1982}
\begin{equation}\label{eq:obf}
{s}(\eta,\tau) = {W_\tau}(\tau) \sigma (\eta,\tau) \otimes h(\eta,\tau) + n_0(\eta,\tau)
\end{equation}
where $\eta,\tau$ are respectively the azimuth and range time, $n_0$ denotes the additive noise, $h(\eta,\tau)$ is the time-variant convolution kernel which can be composed as:
\begin{equation}\label{eq:kncomp}
h(\eta,\tau)=h_1(\eta,\tau)\otimes h_2(\eta,\tau)
\end{equation}
In (\ref{eq:kncomp}), $h_1(\eta,\tau)$ is the two-dimensional azimuth modulation which is responsible for the along-track observation while $h_2(\eta,\tau)$ is range convolution kernel that is identical to the transmitted pulses.

Further, we can sample the continuous-time analog echo $s(\eta,\tau)$ and discrete the reflectivity map $\sigma(\eta,\tau)$, into two-dimensional arrays ${\bf Y}\in \mathbb{C}^{n_\eta^{'}\times n_\tau^{'}}$ and ${\bf X}\in \mathbb{C}^{n_\eta\times n_\tau}$. And then we obtain the following observation model for the strip mode SAR:
\begin{equation}\label{eq:ob2d}
{\bf y}= \textbf{H} {\bf x}+{\bf n}_0
\end{equation}
where ${\bf y}={\rm vec}({\bf Y})\in \mathbb{C}^{l\times 1},l=n_\eta^{'}\times n_\tau^{'}$, ${\bf x}={\rm vec}({\bf X})\in \mathbb{C}^{n\times 1}, n=n_\eta\times n_\tau$, $\bf H$ is the observation matrix acquired from the discrete weight of (\ref{eq:obf}) (more detailed information and construction of the observation model can be seen in \cite{Wu1982}\cite{Wei2010}), and ${\bf n}_0$ is the noise.

\subsection{Formulation of CS-SAR models}
In a CS-SAR model, the data $\bf y$ is sampled and compressed with a proper sampling matrix ${\bf \Theta} \in \mathbb{R}^{m\times l}, m \ll n$, resulting in
\begin{equation}\label{eq:ob2d}
{\bf y}_s={\bf \Theta} \textbf{H} {\bf x}+{\bf n}_s
\end{equation}

When $\bf x$ is a sparse signal, say, most of the entries of $\bf x$ are zeros, the theory of CS tells when and how it can be recovered from the above undetermined linear system with fewer measurements than Nyquist criterion requires\cite{Candes2006cs}\cite{Donoho2006}. Generally, considering an ill-posed linear system ${\bf y}_s={\bf A}{\bf x}$ (${\bf A}={\bf \Theta} {\bf H}$) where $\bf x$ is sparse enough, if the sensing matrix $\bf A$ satisfies some conditions like RIP\cite{Candes2006}, $\bf x$ can be exactly recovered from $\bf y_s$ with the $L_q$ (quasi-norm) ($0\leq q\leq 1$) optimization:
\begin{equation}\label{eq:CSq}
\min\limits_{\bf x}{\kern 5pt}\|{\bf x}\|_q {\kern 5pt}s.t.{\kern 5pt}{\bf y}_s = \textbf{A} {\bf x}
\end{equation}
To solve (\ref{eq:CSq}), we usually use an equivalent regularization scheme with the following optimization problem
\begin{equation}\label{eq:CSregu}
\min\limits_{\bf x}{\kern 5pt}\{\|{\bf y}_s -\textbf{A} {\bf x}\|^2_2+\lambda\|{\bf x}\|_{q}^{q}\}
\end{equation}
where $\lambda$ is a regularization parameter.  The optimization can be efficiently solved by iterative thresholding algorithm (ITA)\cite{Daubechies2004,Xu2010,Blumensath2009}. In detail, an ITA generates a sequence of approximates according to:
\begin{equation}\label{eq:th}
{\bf x}^{(i+1)}={\rm{E}}_{q,\lambda\mu}({\bf x}^{(i)}+\mu \textbf{A}^{\rm H}({\bf y}_s- \textbf{A} {\bf x}^{(i)}))
\end{equation}
where $\mu$ is a normalized parameter which controlls the convergence of the iteration. In (\ref{eq:th}), ${\rm E}_{q,\sigma}$ ($\sigma=\lambda\mu$) is a so-called thresholding operator which is componentwisely defined by
\begin{equation}\label{eq:trdc}
{\rm{E}}_{q,\sigma}({\bf x})=(e_{q,\sigma}({\bf x}_1),e_{q,\sigma}({\bf x}_2),...e_{q,\sigma}({\bf x}_n))^T
\end{equation}
where $e_{q,\sigma}$ can be analytically specified when $q=0,\frac{1}{2},\frac{2}{3},1$. For example, the widely used soft-thresholding, which corresponds to $q=1$, is
\begin{equation}\label{eq:L12}
{e_{1,\sigma }}(x) = \left\{ {\begin{array}{*{20}{c}}
   {{\rm sgn}(x)(\left| x \right|-\sigma),{\kern 12pt} if\left| x \right| \ge \sigma}   \\
   {{\kern 42pt}  0,{\kern 42pt}  otherwise}  \\
\end{array}} \right.
\end{equation}
The iteration (\ref{eq:th}) with (\ref{eq:trdc}) is the fundamental procedure we suggest to use for the CS-SAR imaging.

It can be seen that the main computation load in implementation of (\ref{eq:th}) comes from the calculation of time domain correlation ${\bf A}^{\rm H}{\bf A}{\bf x}$. From the viewpoint of SAR signal processing, this corresponds to the backscattered projection procedure, which is known to be inefficiency for reconstruction, even implemented by convolution as in (\ref{eq:obf}). On the other hand, we notice that there exists efficient focusing methods using MF in traditional SAR signal processing. This type of processing is in the frequency domain, which is much faster. Moreover, unlike the MF based method which is usually decoupled, the sensing matrix $\bf H$ in (\ref{eq:th}), owns intrinsically two-dimensional structure that has to be collected and stored before imaging. Although some compression can be incorporated according to the structure of the matrix, it still consumes a huge memory. All these difficulties then hamper effective applications of the known CS-SAR imaging.

The aim of the present research is to suggest a new CS-SAR imaging method, which replaces the use of the exact observation model by an well-defined approximation, and then makes it possible to reconstruct the sparse scene $\bf x$ via a sequence of 1-D operations. Thus, the very high cost of calculation and memory of the existing CS-SAR imaging methods can be significantly reduced.

\section{Approximated Observation}
In this section, we first explain why an approximated observation is needed and feasible, and then, we provide an example to show how an approximated observation operator can be explicitly constructed by virtue of a concrete example from the inverse of Range-Doppler Algorithm (RDA). A relation between the constructed approximation and the corresponding focus method is analyzed, which then serves as the basis of the development of new method in the next section.


\subsection{Why approximation needed}

It is known that MF is fast with $\mathcal{O}(n\log n)$ complexity, which is, among the others, mainly due to frequency domain operations. More precisely, if we denote $\bf M$ the imaging procedure by MF, like RDA, the SAR raw data can be well focused in some conditions by
\begin{equation}\label{eq:focus}
\tilde {\bf x}={\bf M}{\bf y}
\end{equation}
where $\bf M$ is the traditional MF imaging procedure that can be calculated through decoupling it into a series of 1-D operators in the frequency domain. This normally leads to an $\mathcal{O}(n\log n)$ complexity when fast fourier transformation (FFT) type operations are employed.

Observing these advantages, the purpose of this paper is to accelerate the known CS-SAR imaging procedures, so as to achieve a comparable (at the same order) complexity with the traditional MF based methods.

A natural consideration is then to look for the possibility of integrating CS and MF. However, a direct application on the decoupling of $\bf H$ is impossible, because $\bf H$ intrinsically possesses 2-D structure. Nevertheless, it is known from (\ref{eq:focus}) that $\tilde {\bf x}$ always approximates $\bf x$, say, ${\bf M}{\bf H}\approx {\bf I}$, and hence, ${\bf M}^{-1}$, whenever exists, approximates the observation $\bf H$. In viewing that $\bf M$ is decoupled, it can then naturally be expected that ${\bf M}^{-1}$ is decoupled, too. Thus, we can expect that under certain conditions, some types of approximations of $\bf H$ can be decoupled, so as to bring an $\mathcal{O}(n\log n)$ complexity.

This is why we would like to approximate the observation, and in the following, we will introduce the details on how to construct and what constraints an appropriate approximation observation.

\subsection{How to construct an admissible approximated observation}
\begin{figure}[t]
\centering
\includegraphics[width=0.7\textwidth]{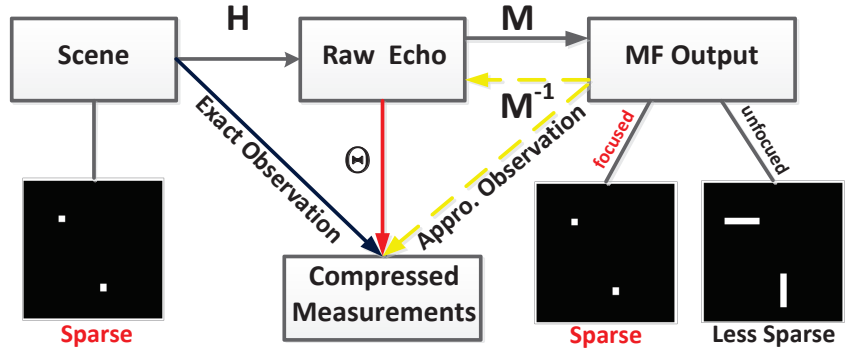}
\caption{The relations between exact observation and approximated observation.}
\label{fig_APOB}
\end{figure}
Fig. \ref{fig_APOB} draws the main relations between CS-SAR observation and MF reconstructions. It can be seen that whenever the imaging procedure $\bf M$ is accurate enough, ${\bf M}^{-1}$ can be viewed as an admissible substitute of $\bf H$. This provides a general principle of how the observation $\bf H$ can be remodeled and approximated by any high precision imaging (or reconstruction) procedure. We formalize this principle further as
\begin{equation}
{\bf G}={\bf M}^{-1}\approx{\bf H}
\end{equation}
where $\bf G$ is any generalized right inverse of $\bf M$, and $\bf M$ is any a high precision imaging procedure. We call $\bf G$ henceforth an approximated observation.

However, since there are many well known imaging procedures that provided various tradeoffs on imaging accuracy and complexity. We need therefore further to define the extent of accuracy and identify the constraints under the CS-SAR framework. To see this, let us compare the exact observation model and the approximated observation model
\begin{equation}
{\bf y}_s={\bf \Theta}{\bf H}{\bf x}={\bf \Theta}{\bf G}\tilde {\bf x}
\end{equation}
It can be seen that by using approximated observation, other than reconstruct $\bf x$, we actually reconstruct $\tilde {\bf x}$ instead, which  is assumed to be an approximation of $\bf x$, when it obeys to the following relation
\begin{equation}\label{eq:apobs}
\tilde {\bf x}={\bf p}_{\epsilon}\circ{\bf x}+{\bf s}_\epsilon
\end{equation}
where $\circ$ denotes the Hadamard product, $\bf p_{\epsilon}$ denote the phase error while ${\bf s}_\epsilon$ is the error for side lobe, or more severely, the artifacts from unfocusing. Formally, when (\ref{eq:apobs}) holds, there exists an acceptable solution with the approximated observation model. However, to find it under CS, we should further emphasis on a better focusing ability of $\bf M$.

As we know, a key parameter in CS-SAR, different from traditional SAR, is the sampling rate which measures how a SAR system benefits from CS. The least amount of samples to ensure the reconstruction is incoherently determined by the sparsity of scene $\bf x$, usually irrelevant of the distribution and phases of targets in the scene. Thus, the difference of the sparsity between $\bf x$ and $\tilde {\bf x}$ determines when and how much the additional measurements does the approximated observation based CS-SAR methods need. It can be immediately seen from (\ref{eq:apobs}) that the difference is uniquely characterized via ${\bf s}_\epsilon$. That is to say, whenever the side lobes reconstructed from $\bf M$ is low enough, ${\bf s}_\epsilon$ can be ignored, then $\bf x$ and $\tilde {\bf x}$ can keep the same sparsity. In this situation, the required least sampling rate of the approximated observation model equals to the original model. In turn, to prevent the approximated observation based CS-SAR method from demanding more samples, the side lobe should be as low as possible, or equivalently to say, a well focusing capacity should be a criterion to determine whether a specific focusing method can be used to construct the approximated observation.

The above discussions tell a fact that the construction of the approximated observation is quite reflexible, which can be acquired straightforwardly based on well established algorithms with additional requirement on the focusing ability. In the next subsection, we present a concrete example using Range-Doppler-algorithm (RDA)\cite{Wu1976} to show how an admissible approximated observation $\bf G$ can be simply constructed based on this principle.

\subsection{A concrete example}
RDA is a very popular procedure for stripmap mode SAR imaging that is simple both in comprehension and in implementation. The procedure (under the low squint case) consists of three main steps (operations): the range compression, RCMC and azimuth compression. In a compact form, the imaging procedure $\rm M$, operated on 2-D array, can then be expressed in the following
\begin{equation}\label{eq:RDA}
\tilde {\bf{X}} = {{\rm{M}}}({\bf{Y}}) = {\bf{F}}_\eta ^{\rm H}\{ {{\bf{P}}_\eta } \circ \rm{C}\langle {{\bf{F}}_\eta }[{{\bf{P}}_\tau } \circ ({\bf{Y}}{{\bf{F}}_\tau })]{\bf{F}}_\tau ^H\rangle \}
\end{equation}
where $\tilde {\textbf{X}}$ ($\tilde {\bf x}={\rm vec}(\tilde {\textbf{X}})$) is the reconstructed 2D SAR image, ${\bf F},{\bf F}^{\rm H}$ respectively are the DFT matrix and inverse DFT matrix (in practice, they are implemented by FFT) to perform, the subscript $\eta,\tau$ denotes the direction of azimuth and range where the FFT performs along, $\textbf{P}_\eta$ and $\textbf{P}_\tau$ are the frequency domain matched filter operations along azimuth and range, which can be always defined respectively by
\begin{equation}\label{eq:amf}
\textbf{P}_\eta(f_\eta;\tau)=\exp[-j\pi/ K_af_\eta^2]{\kern 10pt}\textbf{P}_\tau(f_\tau;\eta)=\exp[-j\pi/ K_rf_\tau^2]
\end{equation}
In (\ref{eq:amf}), $f_\eta,f_\tau$ are the frequency along Doppler and range, $K_a$ and $K_r$ are the azimuth FM rate and the pulse FM rate. In (\ref{eq:RDA}), ${\rm{C}}$ is the RCMC interpolation operator which is essentially a space-variant shift, and always approximated by the truncated sinc-kernel interpolation with ${\bf U}={\rm C}({\bf V})$ as
\begin{equation}\label{eq:RCMCs}
{\bf U}(f_\eta ,\tau ) = \sum\limits_{\tilde \tau } {\bf V}({\tilde f}_\eta ,{\tilde \tau} ){\rm sinc}({\tilde \tau}  - (\tau  + \Delta r(f_\eta ,\tau ))
\end{equation}
where $\Delta r$ is the migration (measurement in time) to be corrected, and $\textbf{U},{\textbf{V}}$ are the signals before and after RCMC, respectively.

With the so specific operations in RDA procedure, we now can derive the inverse of $\bf M$ quite simply by taking the inverse of every sub-procedures. The details are as follows:

\emph{i)} The inverse of Fourier transformations ${\bf F},{\bf F}^{\rm H}$ are known as the inverse transformations, which are given by ${\bf F}^{\rm H},{\bf F}$. It is important to keep the throwaway consistent between the pairs.

\emph{ii)} It is known that phase multiplication is a unitary transformation, so that the inverse is the multiplication of the conjugate phase, ${\bf P}_\eta^\ast,{\bf P}_\tau^\ast$, and the Hadamard multiplication can still be applied in order.

\emph{iii)} The inverse of $\rm C$ is difficult to achieve directly. In fact, $\rm C$ is approximated from the accurate RCMC defined in continuous range time domain. Because the trajectories of targets with different range gates are disjoint, this shift is a one-to-one mapping and the inverse of the origin RCMC exists. We can also approximate through interpolation ${\bf V}=\rm D({\bf U})$ that
\begin{equation}
{\textbf{V}}({{\tilde f}_\eta },\tilde \tau ) =\sum\limits_\tau  {\textbf{U}({f_\eta },\tau ){\rm sinc}(\tilde \tau  - (\tau  + \Delta r({f_\eta },\tau )))}
\end{equation}

Based on the above exposition, the approximated observation ${\rm G}$ deduced from RDA can then be explicitly expressed by:
\begin{equation}\label{eq:apob}
{{\rm G}}({\bf{X}}) = \{ {\bf{P}}_\tau ^* \circ \langle {\bf{F}}_\eta ^{\rm H} \rm{D}[{\bf{P}}_\eta ^* \circ ({{\bf{F}}_\eta }{\bf{X}})]{{\bf{F}}_\tau }\rangle \} {\bf{F}}_\tau ^H
\end{equation}

We show that the so constructed approximated observation $\bf G$ has an interesting property: It is still a linear operator and its conjugate transposition equals to $\bf M$\footnote{It is to say $\bf G$ is nearly unitary, since only a minor approximation on calculation of the inverse is included}.

\textbf{Theorem 1.} ${\bf G}$ is a linear operator with the property ${\bf G}^{\rm H}={\bf M}$.
\begin{IEEEproof}
The linearity of ${\rm G}$ and ${\rm M}$ is obvious because all the sub-operations are linear. Let $\bf x$ denote the vector form of $\bf X$,  namely, ${\bf x}={\rm vec}({\bf X})$. Then, by definition, the linear operators $\rm G$ and $\rm M$ can be written as matrices, and we then have
\begin{equation}\label{eq:G}
{\rm vec}({\rm G}({\bf X})) = {{\textbf{G}}}{\bf x} = {{\bf \hat F}_\tau ^H {\bf \hat P}_\tau ^ * {\bf \hat F}_\eta ^H{{{\bf \hat F}}_\tau }{\bf \hat D}{\bf \hat P}_\eta ^ * {{{\bf \hat F}}_\eta }}{\bf x}
\end{equation}
\begin{equation}\label{eq:M}
{\rm vec}({\rm M}({\bf Y})) = {\textbf{M}}y = {\bf \hat F} _\tau ^H {\bf \hat P} _\eta {\bf \hat C} {\bf \hat F} _\eta {\bf \hat F} _\tau ^H{\bf \hat P} _\tau {\bf \hat F} _\tau {\bf x}
\end{equation}
where
\begin{equation}
{\hat {\textbf F}}_\eta  = {\textbf{I}_{n_{\tau}}} \otimes {{\bf{F}}_\eta },{{\hat {\bf F}}_\tau } = {\bf{F}}_\tau ^T \otimes {{\bf I_{n_\eta}}}
\end{equation}
\begin{equation}
{{\hat {\bf P}}_\eta } = {\rm diag}({\rm vec}({{\bf P}_\eta })),{{\hat {\bf P}}_\tau } = {\rm diag} ({\rm vec}({{\bf P}_\tau }))
\end{equation}
$\hat {\bf C}$ and $\hat {\bf D}$ are real matrices defined by
\begin{equation}\label{eq:CD}
\left\{ {\begin{array}{*{20}{c}}
{\hat {\bf{C}}(i,j) = {\rm{sinc}}(\frac{j - i}{n_\eta } + \Delta {\bf r}(i)/f_s)}  \\
{\hat {\bf{D}}(i,j) = {\rm{sinc}}(\frac{i - j}{n_\eta } + \Delta {\bf r}(j)/f_s)}  \\
\end{array}} \right.{\kern 1pt} {\kern 1pt} {\kern 1pt} {\kern 1pt} {\kern 1pt} {\kern 1pt} if{\kern 1pt} {\kern 1pt} {\kern 1pt} (i - j)\bmod {n_\eta } \equiv 0
\end{equation}
In (\ref{eq:CD}), $\Delta r={\rm vec}(\Delta {\bf R})$, $\Delta{\bf R}$ is the discretion of ${\Delta r}$. Observing from (\ref{eq:CD}) and $f_s$ is the pulse sampling interval, it is easy to check that $\hat {\bf D}=\hat {\bf C}^{\rm T}$. Consequently, comparing (\ref{eq:M}) and (\ref{eq:G}), we conclude that ${{\textbf{G}}}={{\textbf{M}}^{\rm H}}$.
\end{IEEEproof}
Theorem 1 shows that we have actually taken the conjugate transposition of $\bf M$ as an approximated observation of $\bf H$. Such coincidence plays an important role in the new method to be suggested in the next section.

\subsection{Generalization}
The approach we have applied to define the approximated observation by the inverse of RDA procedure can be generalized in the following two folds:

1) For high squint cases, we can incorporate secondary compression in RDA or derive an approximated observation from the inverse of other focusing methods like Chirp-Scaling Algorithm (CSA), $\omega-{\rm k}$ algorithms, so as to enhance the focusing ability. With similar sub-operations, like FFT, phase multiplication and interpolation, the acquisition of the inverse is just as same as RDA.

This principle can be applied to yield more general algorithms, however, we will not enumerate all possible extensions but remind that the decoupled structure of MF makes the inverse always achievable. This is the reason why MF is fast and why we propose to apply the approximated observation instead of the exact observation in the CS-SAR imaging system.

2) In the above derivation, we have assumed that the transmitted signal form is standard chirp and $\bf M$ focus both the azimuth and range direction. In fact, the azimuth modulation is the exclusive property and main difficult of SAR signal processing while the range convolution, which possesses a simple 1-D structure, can be modeled directly. So, we can apply the approximated observation to non-chirp cases, by replacing ${\bf{P}}_\tau ^*$ with the transmitted pulse, which is always recorded in modern SAR systems. This extension is also of specifical necessity in CS-SAR because the design of the pulse form  is also a very important issue.

\section{CS-SAR imaging based on approximated observation}
In this section, we formulate the new CS-SAR imaging method based on the use of approximated observation. An $L_q$ regularization model together with the fast iterative thresholding algorithm will be suggested.

\subsection{The new CS-SAR method}
By replacing the exact observation $\bf H$ using the approximated observation $\rm G$ in (\ref{eq:apob}), we can acquire the following CS-SAR model:
\begin{equation}\label{eq:model}
\min_\textbf{X} \{ \|\textbf{Y}_s-{\bf \Theta}_\eta {\rm G}(\textbf{X}){\bf \Theta}_\tau\|_F^2+\lambda\|\textbf{X}\|^{q}_{q}\}
\end{equation}
where $\|\cdot\|_F$ is the Frobenius norm of a matrix, ${\bf \Theta}_\eta$ and ${\bf \Theta}_\tau$ are the sampling operators in azimuth and range directions, which corresponds to the general sampling operator $\bf \Theta$ in (\ref{eq:ob2d}). It is well defined because the azimuth signal is of discrete form and the range signal is of continuous form, and thus the sampling procedure of the two types of signals are usually physically separated\footnote{Although the sampling scheme on range may vary pulse-by-pulse, we still use this expression which is easily understood}.


Then, due to the linearity of $\rm G$, the model (\ref{eq:model}) can still be very fast solved by ITA, which reads in this case that
\begin{equation}\label{eq:th2d}
{\textbf{X}^{(i + 1)}} = {{\rm E}_{1,\lambda \mu }}({\textbf{X}^{(i)}} + \mu {\rm M}({\bf \Theta} _\eta ^{\rm T}({\textbf{Y}_s} - {{\bf \Theta} _\eta }{\rm G}({\textbf{X}^{(i)}}){{\bf \Theta} _\tau }){\bf \Theta} _\tau ^{\rm T}))
\end{equation}
In this paper, we simply select $q=1$ while parameters $\mu,\lambda$ will be preset according to the next subsection.

Fig. \ref{fig_expla} below shows the flow chart of algorithm (\ref{eq:th2d}), which tells that ITA provides an intuitive explanation in terms of SAR signal processing. It is seen that at each iteration, the ITA can be decomposed into mainly three procedures: the compressed data simulation, the matched filter on the residual and the thresholding for new estimation. Physically, this means that in every iteration of the ITA, the useful information in the residue (not the raw data) is first extracted by MF and then added to the current estimate to yield a new update, finally the thresholding procedure enforces the sparsity through regularizing the noise and ambiguity from under-sampling.
\begin{figure}[h]
\centering
\includegraphics[width=0.8\textwidth]{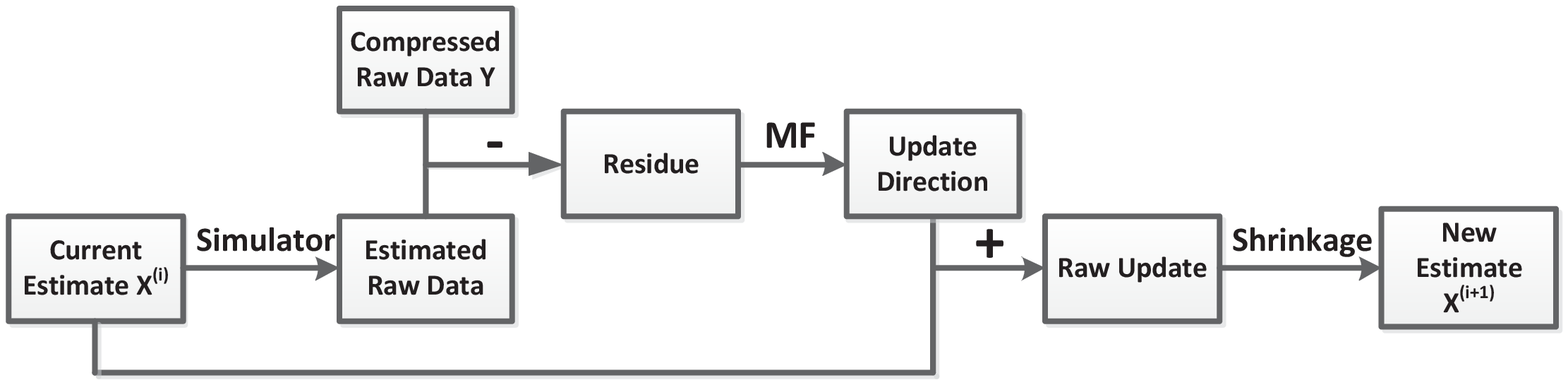}
\caption{An explanation of the proposed algorithm in one-step iteration. The compressed data is alternatively processed by MF and thresholding.}
\label{fig_expla}
\end{figure}

The algorithm stops when converges or achieves the maximum number of iterations. For convenience of use, we list pseudo-code of ITA (\ref{eq:th2d}) as Algorithm 1 below. We further show how the parameters in (\ref{eq:th2d}) can be set adaptively.

\begin{algorithm}
\caption{: Iterative thresholding algorithm for approximate observation based CS-SAR imaging}
\label{alg:ISTA}
\begin{algorithmic}[1]
{\small
\REQUIRE
   SAR raw echoes $\textbf{Y}_s$, approximated observation operator $\rm G$ and $\rm M$, sampling operator ${\bf \Theta}_\eta,{\bf \Theta}_\tau$
\ENSURE
   The recovery image $\textbf{X}^*$
\renewcommand{\algorithmicensure}{\textbf{Initial:}}
\ENSURE
    $\textbf{X}^{(0)}=0$,$\lambda,\mu$ and max iteration $I_{\max}$
\FOR{$i=0$ to $I_{\max}$}
\STATE  Residue: $\tilde {\textbf{R}}^{(i)}=\textbf{Y}_s-{\bf \Theta}_\eta {\rm G}(\textbf{X}^{(i)}){\bf \Theta}_\tau$
\STATE  MF on residue:$\Delta \textbf{X}^{(i)}={\rm M}({\bf \Theta}_\eta^T \textbf{R}^{(i)}{\bf \Theta}_\tau^T)$
\STATE Thresholding: $\textbf{X}^{(i+1)}={\rm E}_{1,\lambda\mu}(\textbf{X}^{(i)}+\mu\Delta \textbf{X}^{(i)})$
\ENDFOR
}
\end{algorithmic}
\end{algorithm}

\subsection{Parameter Setting}
There are two parameters $\mu$ and $\lambda$ in (\ref{eq:th2d}) that need to be set. First, $\mu$ controls convergence of the ITA that the inverse should obeys.
\begin{equation}\label{eq:mure}
0<\mu^{-1}<\|{\bf A}\|^2_2
\end{equation}
However, it is difficult to calculate $\|{\bf \Theta}_\eta{\rm G}{\bf \Theta}_\tau\|_2^2$ directly, where operator $\rm G$ is included. As an alternative, we adopt the adaptive step selection strategy in \cite{Blumensath2010} as:
\begin{equation}\label{eq:mure}
\mu_i=\|{\bf \Theta}_\eta {\rm G}(\Delta{\bf X}_k^{(i)}){\bf \Theta}_\tau\|_F^2/\|\Delta{\bf X}_k^{(i)}\|_F^2.
\end{equation}
where $\Delta{\bf X}_k^{(i)}$ equals to $\Delta{\bf X}^{(i)}$ at the support of ${{\bf X}^{(i-1)}}$, and equal to zero elsewhere. It is easy to demonstrate that $\mu_i$ satisfies (\ref{eq:mure}), and as reported in \cite{Blumensath2010}, such choice has an additional advantage of accelerating the algorithm.

Further, the regularization parameter $\lambda$, which functions to compromise the reconstruction precision and the sparsity of the solutions obtained, has a substantial impact on the imaging result. Fortunately, as a part of the $L_q$ regularization theory, the optimally $\lambda$ has been resolved in \cite{Xu2010}, whenever the problem's sparsity is known. More precisely, assume the considered problem has sparsity $k$ (i.e. a $k$-sparsity problem), then the optimal setting problem of parameter $\lambda^\ast$ is shown to satisfy
\begin{equation}
{\lambda ^*} \in \left[ {\left| {{b_\mu }({{\bf x}^ * })} \right|_{k + 1}/\mu,\left| {{b_\mu }({{\bf x}^ * }} \right|_k^{})}/\mu \right]
\end{equation}
where ${b_\mu}({{\bf x} })={\bf x}+\mu\Delta {\bf x}$, $\left| {{b_\mu }({{\bf x}^ * })} \right|_{k}$ is its $k$-th largest component in magnitude.

Therefore, we suggest the setting that in the $n$th iteration $\lambda_i =\left| {{b_\mu }({{\bf x}^{(i)}})} \right|_{k + 1}/\mu_i$ ( $\lambda_i\mu_i$ is independent of $\mu_i$). The sparsity $k$, which determines $\lambda_i$, can be much flexible to be set, say, based on a prior upper estimation on sparsity of the target scene.

\subsection{Computation Cost}
Let us compare the computational complexity and the memory occupation of the suggested CS-SAR imaging method (\ref{eq:model}), as compared with the known CS-SAR model (\ref{eq:CSregu}). The purpose is to see how much reduction of computational cost of the new model has been brought. In the calculation, we have used some standard notations which are: the number of required iteration $I$, the sampling rate $s$, the number of range gates $n_\eta$, the number of range lines $n_\tau$ ($n=n_\tau\times n_\eta$), the number of samples of sent pulse $u_\tau$, the number of samples of the synthetic aperture time $u_\eta$ (they equal to the time bandwidth product (TBP) in each direction), and the TBP of radar signal $u=u_\eta\times u_\tau$.

With these notations, we can calculate the computational complexity of the approximate observation based CS-SAR, $C_a$, and the computational complexity of the exact observation based CS-SAR, $C_e$ as follows. For $C_a$, it includes calculations of an inverse MF procedure and a MF procedure, which has commonly the computational complexity of $\mathcal{O}(n\log_2 n)$, together with a decoupled thresholding operator with complexity $\mathcal{O}(n)$ in a single step. Thus, the total cost is at the order $C_a=\mathcal{O}(I n\log_2n)$. For $C_e$, it includes calculations of a single iteration, two matrix multiplications and the thresholding procedure. Since there are only few non-zero entries in $\bf H$, say, nearly $us$ in every column, when coding it using two-dimensional convolution,  it needs at least $2uns$ complex multiplication. Thus, we find that the total cost is $C_e=\mathcal{O}(Iuns)$. Then, the ratio between  $C_e$ and $C_o$ is given by $r_C$:
\begin{equation}\label{eq:costr}
r_C=\mathcal{O}(us/\log_2n)
\end{equation}
It is seen from (\ref{eq:costr}) that the ratio $r_C$ depends linearly on the TBP of radar signal $u$. In SAR applications, the $u$ is always designed very large (thousands even millions) to improve the reconstruction signal to noise ratio (SNR), which will bring very high computational cost of the time domain reconstruction method.

The memory loads of the approximate observation based CS-SAR, $M_a$, and the memory loads of the exact observation based CS-SAR, $M_e$ can be estimated in the subsequent way. For $M_a$, it contains only the storage of input, output and several parameter matrices (i.e., azimuth matched filter, range matched filter and the amount of migration in RDA), which is summed up to $\mathcal{O}(n)$ bytes memory occupation. For $M_e$, although no filters are stored, it needs additionally to store a sensing matrix, with the number of non-zero entries of $uns$. But, because the Doppler history with same range cell share the identical patterns, we only need to store an intact holistic pattern (the convolution kernel) for each range gate to achieve a compression, resulting in an additional memory occupation of $16un_\tau$ bytes (a complex number occupies 16 bytes memory), as compared with $M_a$. This additional cost can be very large in spaceborne SAR systems. For example, when $u=10^6$ and $n_\tau=10^4$, it requires more than 100 GB memory to store the array. But the memory cost of RDA is only a few hundreds MB in the same condition. This will further hamper the application of time domain methods into practice.

Finally, the required number of iteration steps is difficult to compare analytically, but in practice, no obvious difference is observed.

\subsection{A Summary}
From the analysis in the previous subsections, we can see that the suggested new model (\ref{eq:model}) and method (\ref{eq:th2d}) have constituted a more efficient CS based SAR imaging method. While preserving CS features, the new method has the following exclusive advantages:

 $\bullet$ \emph{Lower computational cost}: Due to the use of approximated observation, the method only involves 1-D operations which makes the imaging process extremely efficient. It has reduced the computational complexity of the existing exact observation based method  significantly, as shown in (\ref{eq:costr}). Meanwhile, taking full advantages of the decoupled structure after approximating the observation, the proposed method can save the memory cost with a remarkable amount, which is sometimes of more significances.

 $\bullet$ \emph{Higher Compatibility}: As compared with the traditional MF based SAR imaging procedure, the new method uses the same or similar operations, except an additional thresholding operation that yields the sparsity of solution. In particular, the new procedure can be seen as a successive iterative refinement of the well known MF based method,  which makes the new method more consistent. As a result, the proposed model requires little modification of the existing SAR imaging algorithms,  which makes the combination of MF and CS much simpler.

 All these features make the suggested new CS-SAR imaging method more useful and efficient, and particularly possible to be applied in high dimensional SAR applications. We will provide simulations and applications in the next section to further support such benefit.

\section{Simulations and applications}

In this section, several simulations and applications are provided to demonstrate the effectiveness and efficiency of the proposed CS-SAR imaging method. For abbreviations, we denote by CSRDA the CS-SAR imaging method (\ref{eq:model}) with the approximated observation $\rm G$ acquired from the inverse of RDA, and by CSEO the CS-SAR method (\ref{eq:CSregu}) with the exact observation $\bf H$.

We first conduct a series of simulations to compare the performance of the CSRDA method, the CSEO method and the traditional RDA method in terms of reconstruction ability, reconstruction quality and reconstruction cost. Then, we apply the CSRDA to some real SAR imaging tasks from RADARSAT-1, which then further demonstrates the outperformance of the suggested method.

The sampling scheme used in the simulations are specified as follows. In the azimuth direction, we employed random downsampling, realized by selecting random rows from the raw data $\bf Y$, with sampling rate $s_a$. In the range direction, we picked up random samples independently on each sampled echoes in azimuth, with sampling rate $s_r$. And we keep the ratio between $s_a$ and $s_r$ as $1:5$\footnote{The suggested sampling strategy was designed to comprehensively compare the reconstruction algorithms. The proposed model itself is adaptively to more complicated sampling schemes, for example, jitter sampling in the azimuth direction \cite{Patel2010} and random demodulation\cite{Tropp2010} in the range direction. }.

Table 2 lists the primary SAR parameters used in both simulations and applications. All the experiments were conducted on a work station of 8-core 2.4GHz CPU with 32G memory. The CSRDA was implemented in MATLAB 2012a while the CSEO using optimized convolution was implemented in C++ with parallel codes and careful array operations.

\begin{table}[h]
\renewcommand{\arraystretch}{1.3}
\caption{Primary Parameter of SAR system and geometry}
\label{table:para}
\centering
\begin{tabular}{|c||c||c||c|}
\hline
Parameter & Symbol & Simulation & RadarSat-1\\
\hline
Slant range of scene center($\rm km$) &$R_c$ & 20 & 1016.7\\
\hline
Effective radar velocity($\rm m/s$) &$V$ & 350 & 7062\\
\hline
Beam squint angle($\rm rad$) &$\theta$ & 0 & 0.06\\
\hline
Radar center frequency($\rm MHz$) &$f_0$ & 5000 & 5300\\
\hline
Pulse repetition frequency($\rm Hz$) &$F_a$ & 175 & 1256.98\\
\hline
Range FM rate($\rm MHz/\mu s$) &$K_r$ & 37.5 & 0.72135\\
\hline
Pulse duration($\rm \mu s$) &$T_r$ & 2 & 41.75\\
\hline
Sampling rate($\rm MHz$) &$F_r$ & 75 & 32.317\\
\hline
\end{tabular}
\end{table}

\subsection{Simulations}
In the simulations, the scene was taken as $180\times180$ while the scattered coefficients were chosen with unit amplitude and uniform random phases. The raw data were first generated in time-domain by exact slant range and then sampled with different rate, to yield the compressed measurements. The sparsity parameter $k$ was kept the same for CSRDA and CSEO and the maximum iteration steps was set to 100 for both methods. The aims of simulations is then to compare the reconstruction ability (RA), reconstruction quality (RQ) and reconstruction cost (RC), of each competitive SAR imaging methods. These are measured respectively by the lowest amount of measurements a method can successfully  reconstruct an image, the side lobe and resolution of reconstructed point target, and the computation time cost by a method to recover the image.

\begin{figure}[t]
\centering
\subfigure[\ ]{
\begin{tabular}{|c|}
\hline
\mgape{\includegraphics[width=0.2\textwidth]{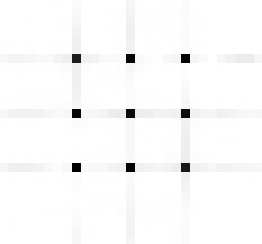}}\\
\hline
\end{tabular}
}
\hfil
\subfigure[\ ]{
\begin{tabular}{|c|}
\hline
\mgape{\includegraphics[width=0.2\textwidth]{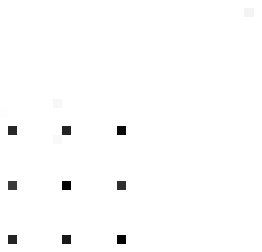}}\\
\hline
\end{tabular}
}
\hfil
\subfigure[\ ]{
\begin{tabular}{|c|}
\hline
\mgape{\includegraphics[width=0.2\textwidth]{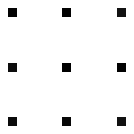}}\\
\hline
\end{tabular}
}
\\
\subfigure[\ ]{
\begin{tabular}{|c|}
\hline
\mgape{\includegraphics[width=0.2\textwidth]{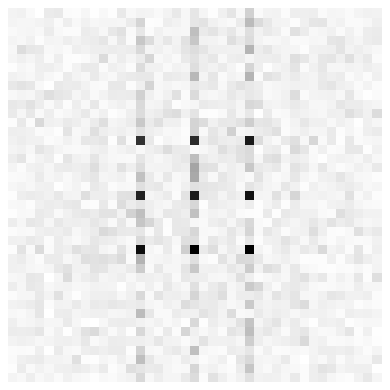}}\\
\hline
\end{tabular}
}
\hfil
\subfigure[\ ]{
\begin{tabular}{|c|}
\hline
\mgape{\includegraphics[width=0.2\textwidth]{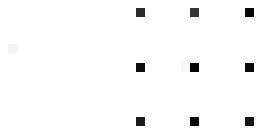}}\\
\hline
\end{tabular}
}
\hfil
\subfigure[\ ]{
\begin{tabular}{|c|}
\hline
\mgape{\includegraphics[width=0.2\textwidth]{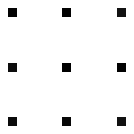}}\\
\hline
\end{tabular}
}
%
\hfil
\caption{The reconstruction results of 9 point targets simulations with different sampling rate. From left to right are the reconstruction results of RDA, CSRDA and CSEO, respectively. And the top row is with full samples while the bottom with 10\% samples}
\label{fig_radars}
\end{figure}

\begin{figure}[t]
\centering
\subfigure[\ ]{
\begin{tabular}{|c|}
\hline
\mgape{\includegraphics[width=0.2\textwidth]{CSRDA207}}\\
\hline
\end{tabular}
}
\hfil
\subfigure[\ ]{
\begin{tabular}{|c|}
\hline
\mgape{\includegraphics[width=0.2\textwidth]{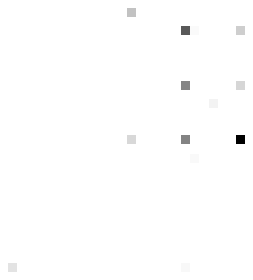}}\\
\hline
\end{tabular}
}
\hfil
\subfigure[\ ]{
\begin{tabular}{|c|}
\hline
\mgape{\includegraphics[width=0.2\textwidth]{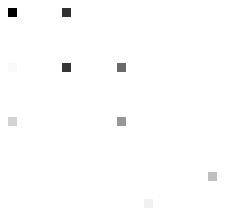}}\\
\hline
\end{tabular}
}
\\
\subfigure[\ ]{
\begin{tabular}{|c|}
\hline
\mgape{\includegraphics[width=0.2\textwidth]{CSOM207}}\\
\hline
\end{tabular}
}
\hfil
\subfigure[\ ]{
\begin{tabular}{|c|}
\hline
\mgape{\includegraphics[width=0.2\textwidth]{CSOM206}}\\
\hline
\end{tabular}
}
\hfil
\subfigure[\ ]{
\begin{tabular}{|c|}
\hline
\mgape{\includegraphics[width=0.2\textwidth]{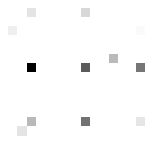}}\\
\hline
\end{tabular}
}
\caption{The detailed comparison on needed least samples to reconstruct the image. The top row is the results from CSRDA, and the bottom row is from CSEO. From left to right are the results corresponding to 0.65\%, 0.55\%, 0.45\% sampling rate, respectively.}
\label{fig_radars1}
\end{figure}

\emph{1) RA comparison: } A set of simulations was made where 9 targets were located at the center with intervals of 6 samples. We varied the sampling rate ranged from 100\% to 0.6\% and added gaussian noise with level of 20 dB. We applied RDA, CSRDA and CSEO to this experiment with sparsity parameter $k=18$. Some of the simulation results are shown in Fig. \ref{fig_radars} and \ref{fig_radars1}.

It is seen from the top row of Fig. \ref{fig_radars} that with full samples (namely, with 100\% sampling rate), all the methods RDA, CSEO and CSRDA can successfully recover the scene, say, the amplitude of the target is maintained and no false target is observed. But the reconstruction of RDA is with serious side lobes, which is not observed in CSRDA and CSEO. This shows the exclusive advantage of the sparse regularization based CS-SAR imaging methodologies, as reported in \cite{Cetin2001}. When we reduce the sampling rate, say, 10\% samples, as seen in the bottom of Fig. \ref{fig_radars} RDA fails to recover the scene, while CSEO and CSRDA both can not only perfectly reconstruct the scene but with significantly reduced side lobes. In this case, no visible difference can be observed for CSEO and CSRDA. Nevertheless, when the sampling rate continues reducing as in Fig. \ref{fig_radars1}, we found that both CSRDA and CSEO can reconstruct the image with only 0.65\% of the samples. However, CSRDA fails with 0.55\% samples while CSEO fails until the sampling rate takes 0.45\%.

All the results consistently show that benefited from sparse regularization, the approximate observation based CS-SAR method can reconstruct sparse scenes with far less samples than Nyquist rate requires. However, caused by approximation, it requires a little more samples to reconstruct the scene.

\begin{figure}[t]
\centering
\subfigure[\ ]{\includegraphics[width=0.4\textwidth]{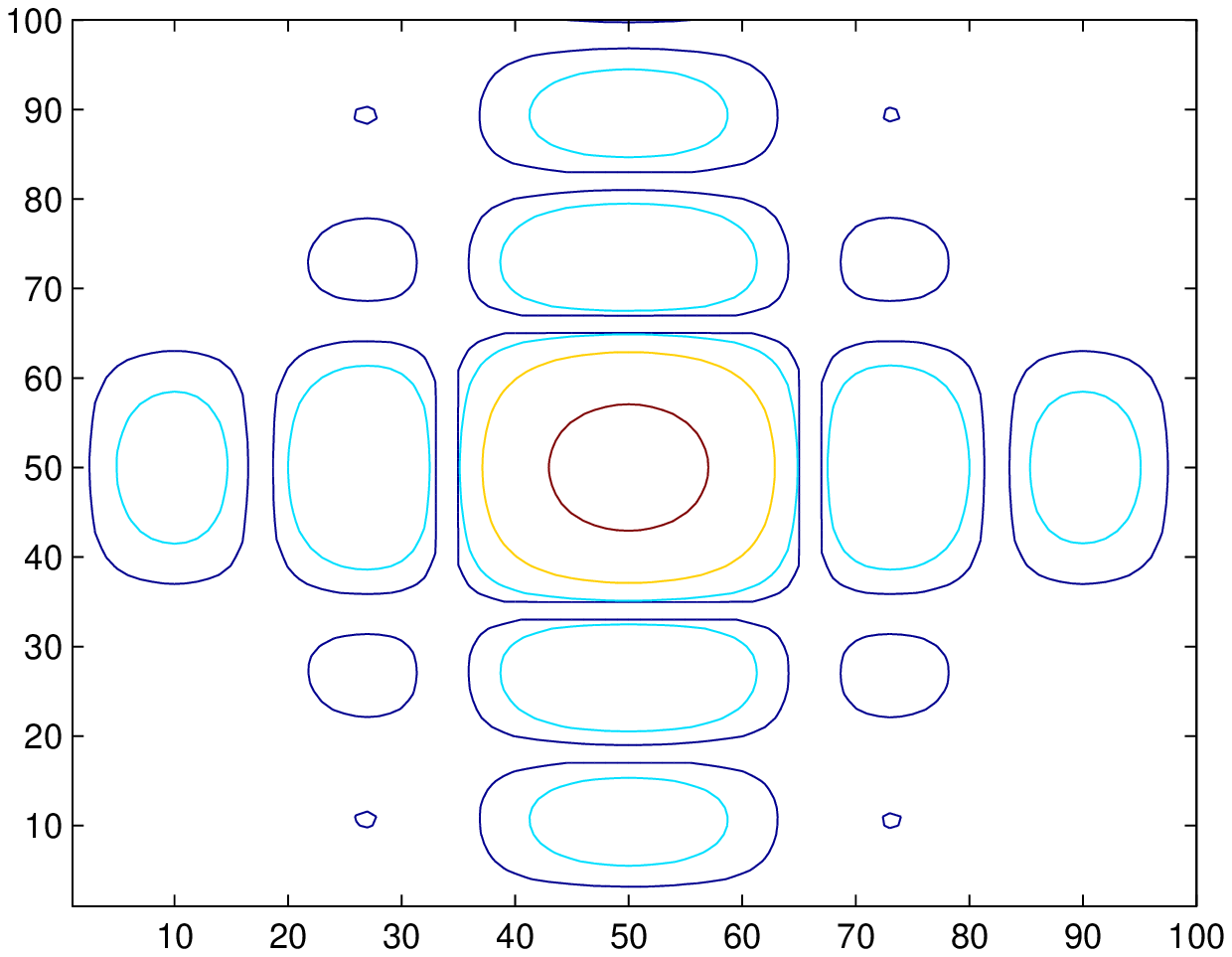}}
\hfil
\subfigure[\ ]{\includegraphics[width=0.4\textwidth]{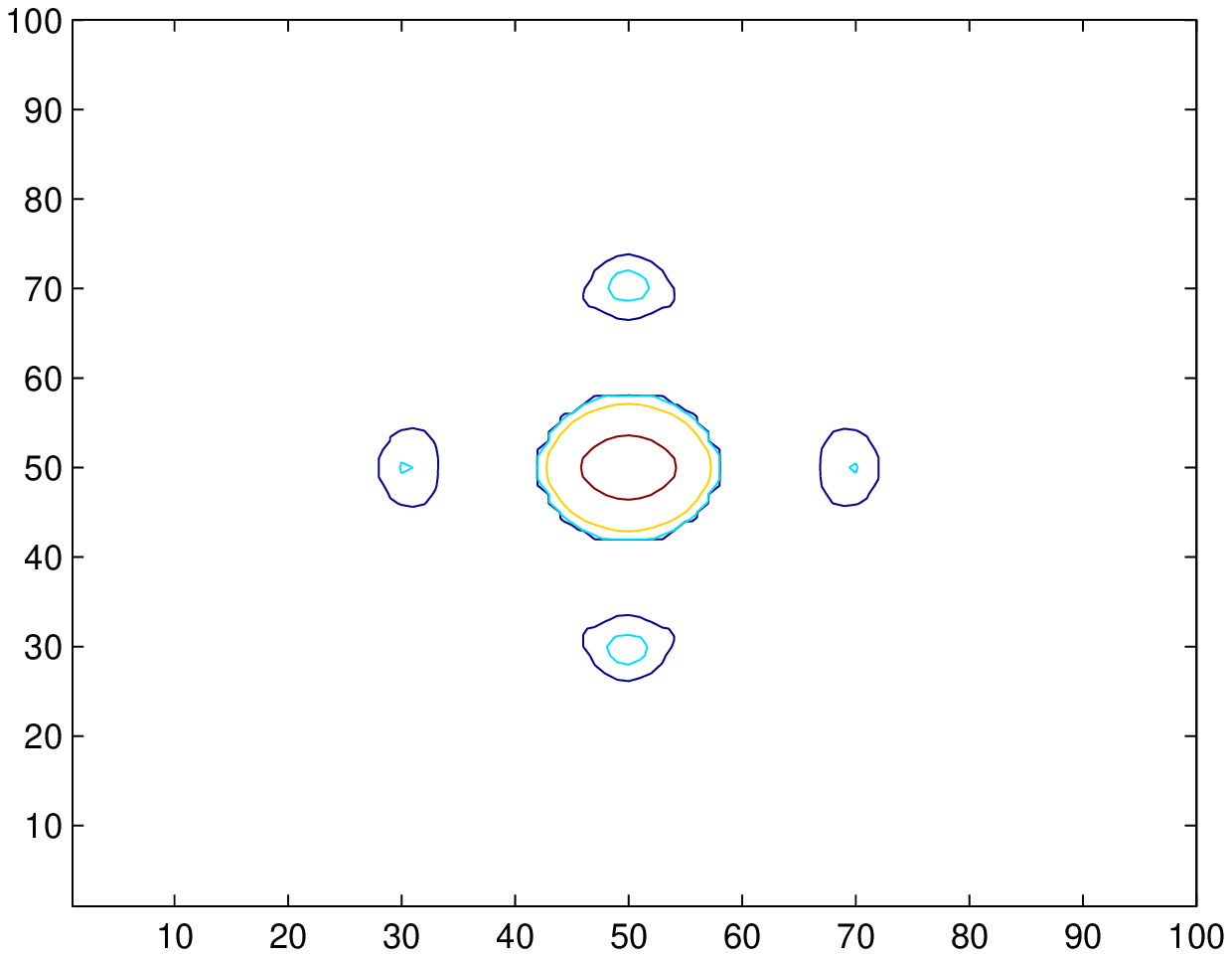}}
\hfil
\begin{tabular}{|c||c||c||c||c|}
\hline
Method & PSLR(R) & PSLR(A) & IRW(R) & IRW(A)\\
\hline
RDA &-13.32 dB & -13.32 dB & 15 samples & 15 samples\\
\hline
CSRDA &-22.71 dB & -21.32 dB & 8 samples & 8 samples\\
\hline
\end{tabular}
\caption{Contours of magnitude. The red, yellow, light blue, blue contour lines are corresponding to the value of -3 dB,-13 dB,-23 dB and -33 dB. (a) Contours from RDA. (b) Contours from CSRDA. (c) Contours from CSEO.}
\label{fig_contour}
\end{figure}

\emph{2) RQ comparison:} Sparse regularization was demonstrated in SAR and CS-SAR imaging the ability of reducing the side lobe and simultaneously improving the resolving ability\cite{Cetin2001}. Hence, we are interested in whether the enhancement is kept when approximated observation is included, especially when the effect of the accuracy of the observation is excluded. To illustrate it, we compare the reconstruction quality of RDA and CSRDA in terms of side lobe and spatial resolution, when successful reconstruction is achieved. The side lobe is evaluated via the peak side lobe ratio (PSLR), defined as the ratio of the peak intensity of the most prominent side lobe to the peak intensity of the main lobe, say, the smaller the PSLR, the better an algorithm. The spatial resolution is measured by the impulse response width (IRW), defined as the width of the main lobe of the impulse response, measured by 3 dB below the peak value, or the minimum distance an algorithm can separate two targets, which should be also as smaller as better. We have perform a one point simulation with the upsampled factor 16 while the target is analyzed by a $16\times 16$ chip centered on the peak, to yield a more detailed analysis on both the main lobe and the side lobe. The sampling rate is fixed (10\% for CSRDA while 100\% for RDA) to ensure a successful reconstruction, and the sparsity parameter $k$ is set 600. Some of the simulation results are given in Fig. \ref{fig_contour}.

Fig. \ref{fig_contour} show the contours of the reconstruction results, with contour lines of -3 dB (the boundary of main lobe) and -13 dB (the PSLR of  traditional MF output). The comparison intuitively shows that both the area of the main lobe and the PSLR in the side lobe reconstructed from CSRDA is much smaller than those reconstructed from RDA. Further in the table, details on the reconstruction quality are presented in azimuth and range directions. It is seen that the width of main lobe reconstructed from RDA is 15 samples both in range and azimuth, but for CSRDA, the width is only 8 samples in the two directions. For the side lobe, the PSLR reconstructed from RDA is -13.32 dB in range and azimuth direction, which is very obvious. But the CSRDA reduces the PSLR to -21.3 dB and -22.7 dB in azimuth and range, respectively.

\begin{figure}[h]
\centering
\subfigure[\ ]{\includegraphics[width=0.4\textwidth]{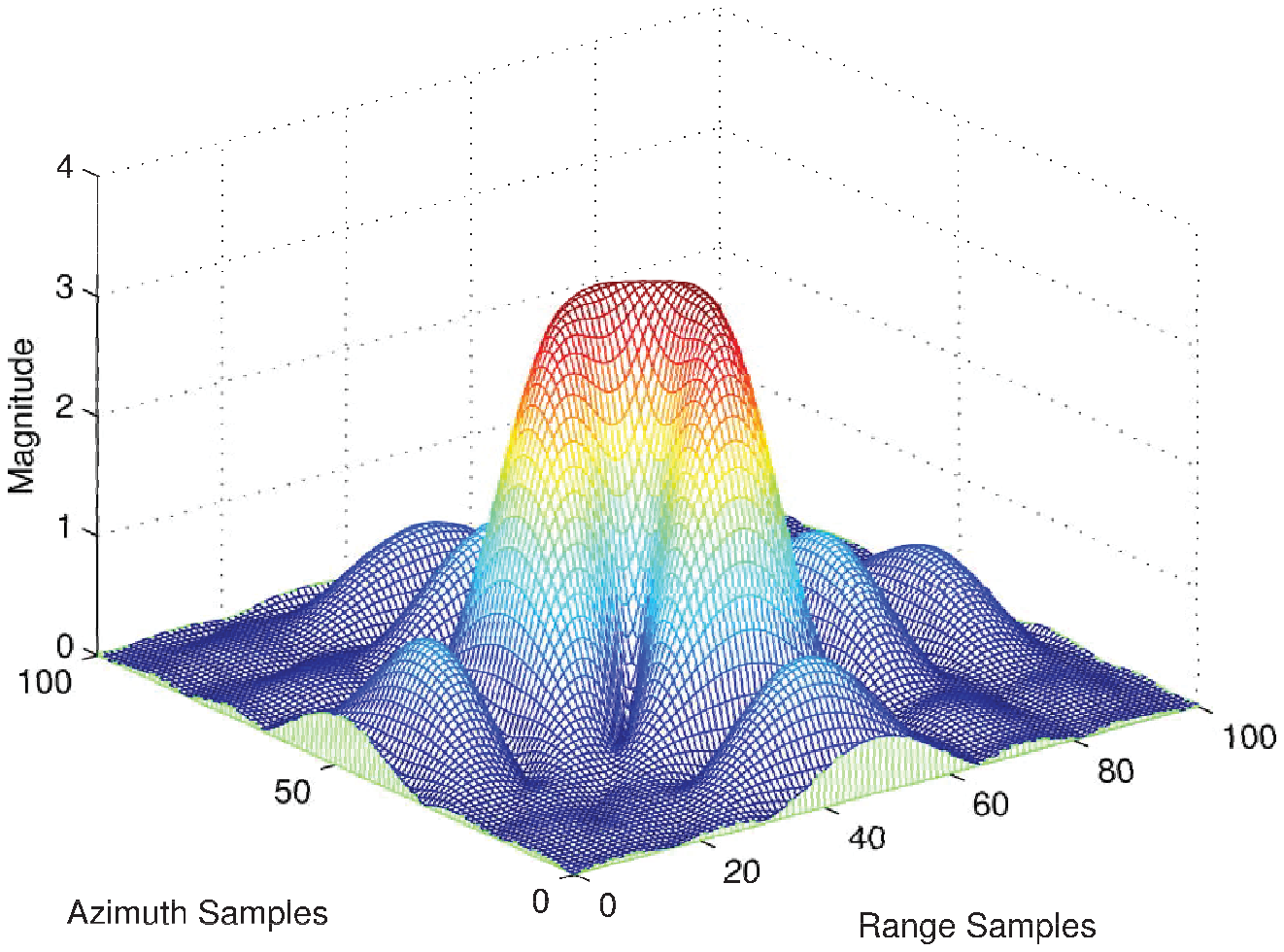}}
\hfil
\subfigure[\ ]{\includegraphics[width=0.4\textwidth]{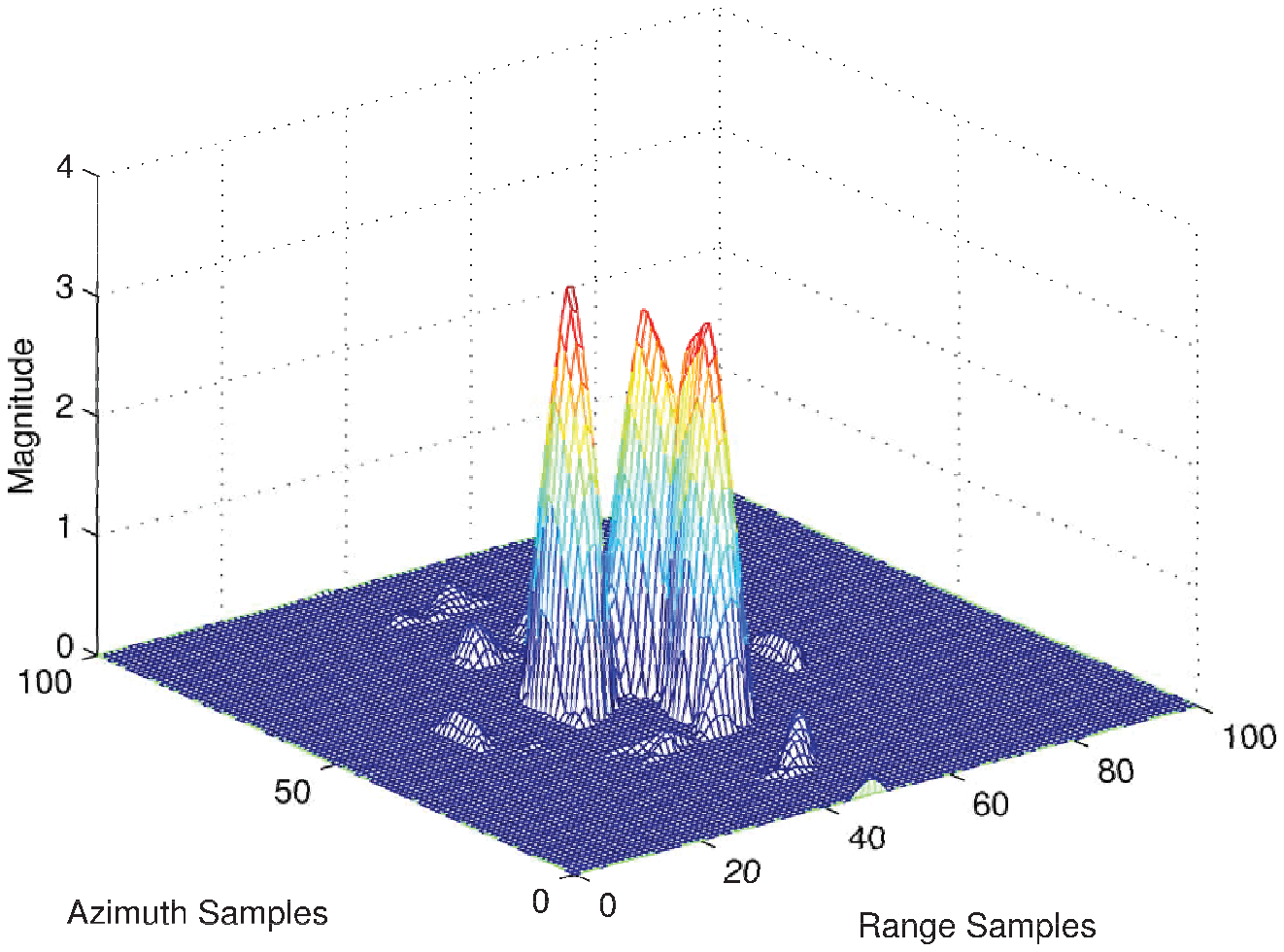}}
\hfil
\caption{Superresolution ability of the proposed method. (a) Result of RDA. (b) Result of CSRDA.}
\label{fig_seperate}
\end{figure}

We further demonstrate the enhancement of resolution by CSRDA in another way. We add two point targets in the above simulation, with respectively azimuth and range interval of 12 samples from the center. We then test whether the three targets can be separated from CSRDA. It is seen from the simulation result (Fig. \ref{fig_seperate}(a)) that the reconstructed result of traditional RDA exhibits an overlapping of the main lobe, thus the targets are inseparable. By using CSRDA in Fig. \ref{fig_seperate}(b), one can clearly distinguish the locations of the three points, demonstrating an obvious enhancement of the resolution.

All of the simulations in this subsection demonstrate that by combining MF with sparse regularization, we can reduce the side lobe and improve the resolution simultaneously to a great extent. Note that, these two goals have been regarded as trade-offs traditionally, if only MF is employed.

\begin{figure}[h]
\centering
\subfigure[\ ]{\includegraphics[width=0.4\textwidth]{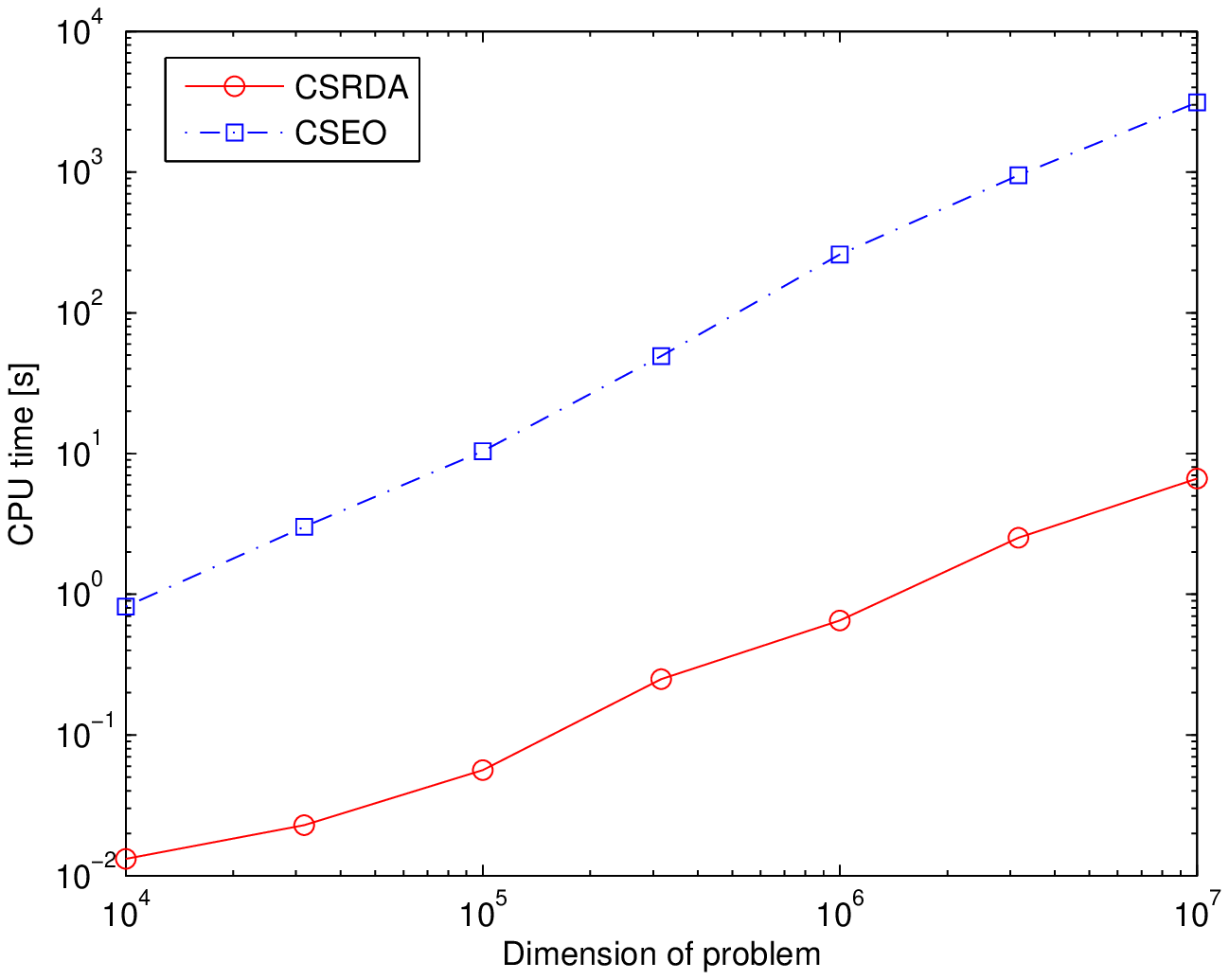}}
\hfil
\subfigure[\ ]{\includegraphics[width=0.4\textwidth]{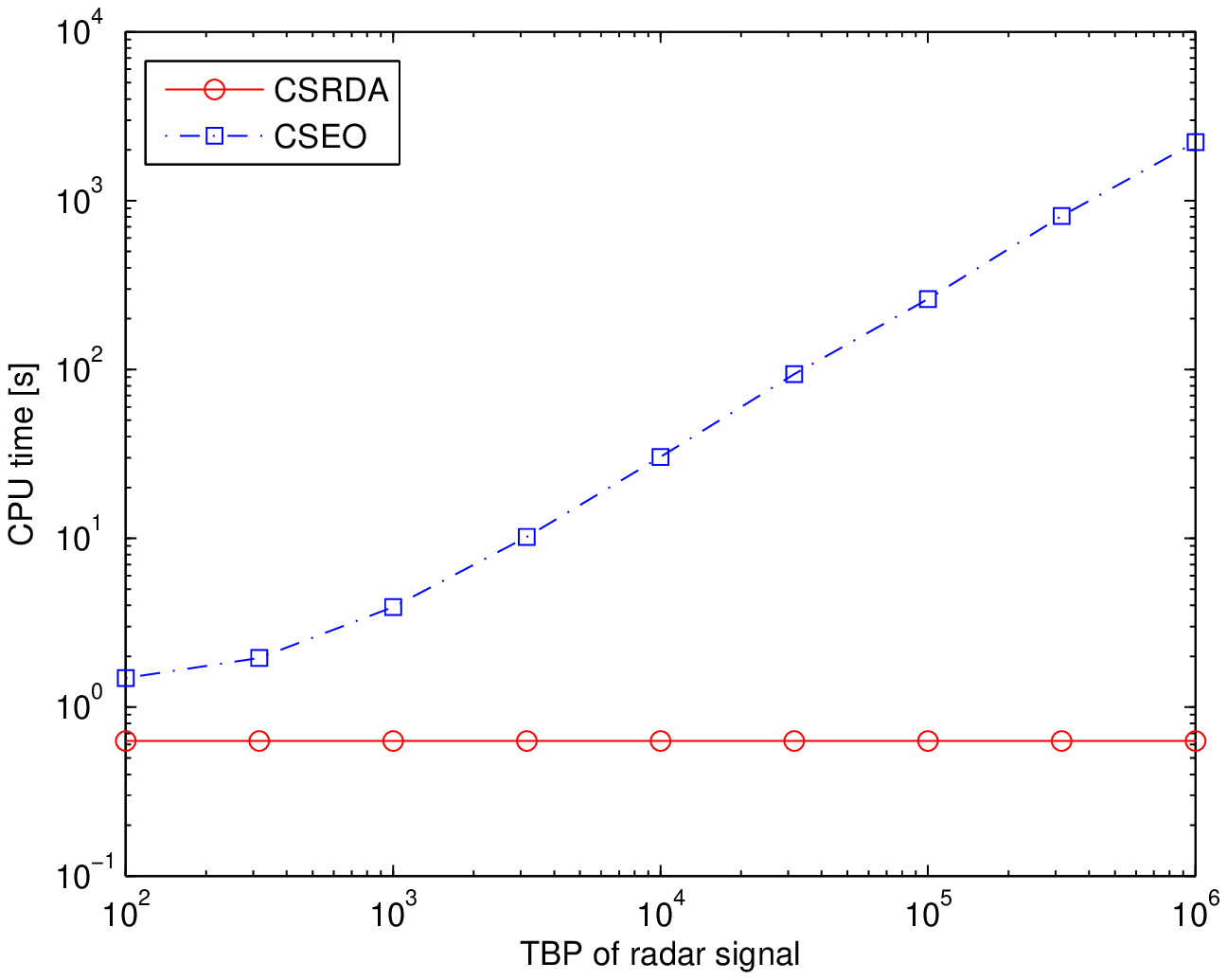}}
\caption{The CPU time of CSRDA and CSEO, where (a) $u$ is fixed as $10^5$. (b) $p$ is fixed as $10^6$. }
\label{fig_RC}
\end{figure}

\emph{3) RC comparison:} Finally, we compare the CPU time takes by CSRDA and CSEO. According to the analysis in section IV, the computational complexity of CSRDA and CSEO depends on the scene size $n$ and TBP of radar signal $u$. So we generated 10 examples for each set of fixed scene size $n$ and TBP of radar signal $u$ in simulation, with a constant sampling rate $s$ 10\%. The average computational time in a single iteration of the two methods was then recorded. The comparison results are shown in Fig. \ref{fig_RC}.

As we can observe from Fig. \ref{fig_RC}, when $u$ is fixed as $10^5$, the CSRDA scales very well to very high dimensional problems, since even with $p=10^7$, it only takes several second to finish an iteration. The CSEO is also insensitive to dimension when $u$ is fixed. However, the CPU time of CSEO is consistently higher than that of CSRDA, with a ratio around 100. On the other hand, it is seen from Fig. \ref{fig_RC}(a) that when $n$ is fixed to $10^7$, the CPU time of CSRDA is constantly 3s, but, the CPU time of CSEO depends linear on $u$, which becomes more and more costly as $u$ increases.

This RC comparison shows that the approximate observation based CSRDA is much faster than the time domain method CSEO, benefited from the $\mathcal{O}(n\log _2 n)$ computational cost of MF. And the computational complexity of CSEO depends linearly on both the TBP of radar pulse and the scene size. So when the $u$ is large, i.e. $10^6$ which is commonly in spaceborne cases, CSRDA is expected to accelerate the CS-SAR reconstruction more than thousands of times. This acceleration of computational time together with the shown memory saving capacity demonstrate the superiority of the proposed method.
\begin{figure}[t]
\centering
\subfigure[RDA(100\%) ]{\includegraphics[width=0.22\textwidth]{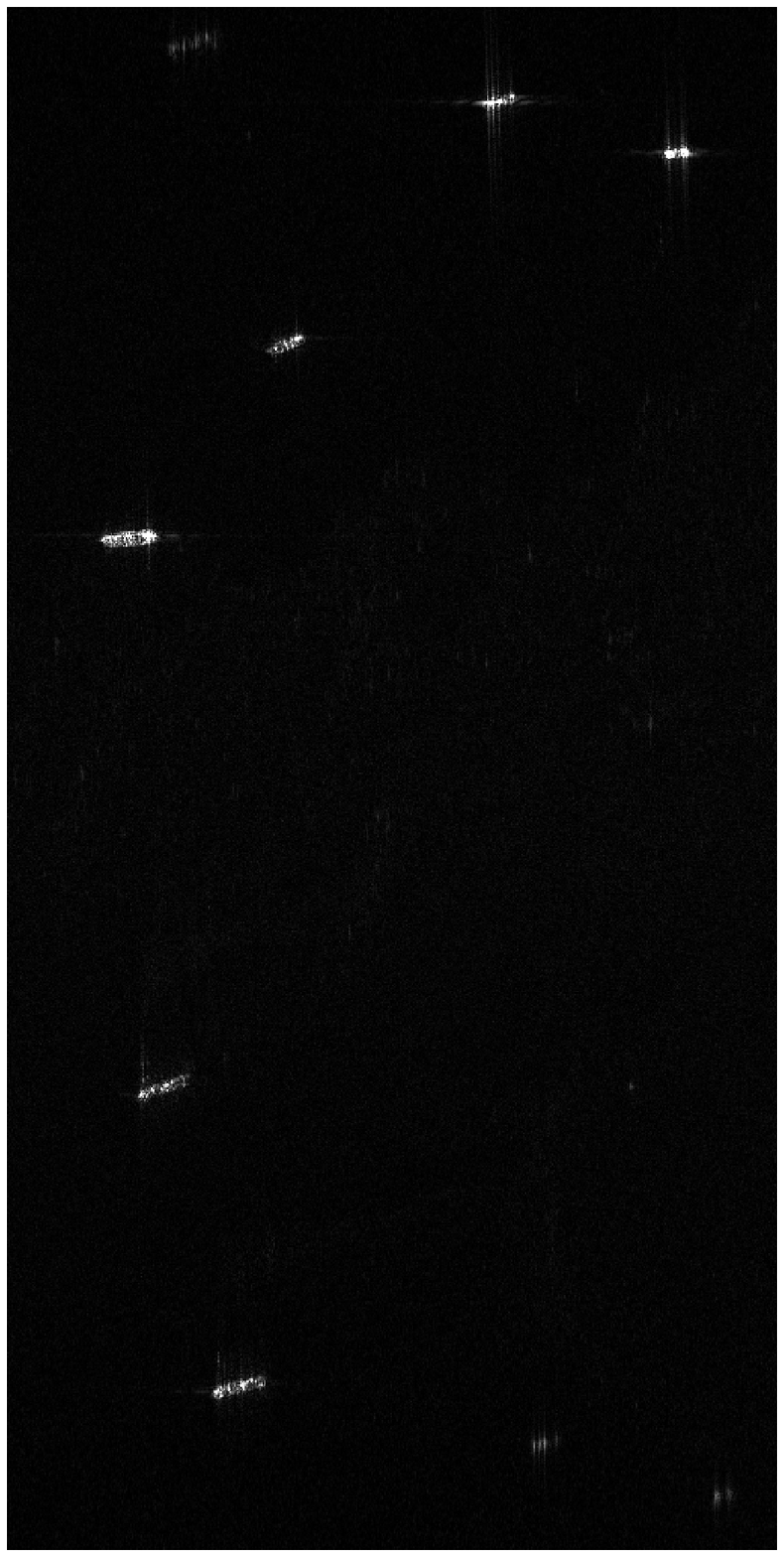}}
\hfil
\subfigure[CSRDA(20\%) ]{\includegraphics[width=0.22\textwidth]{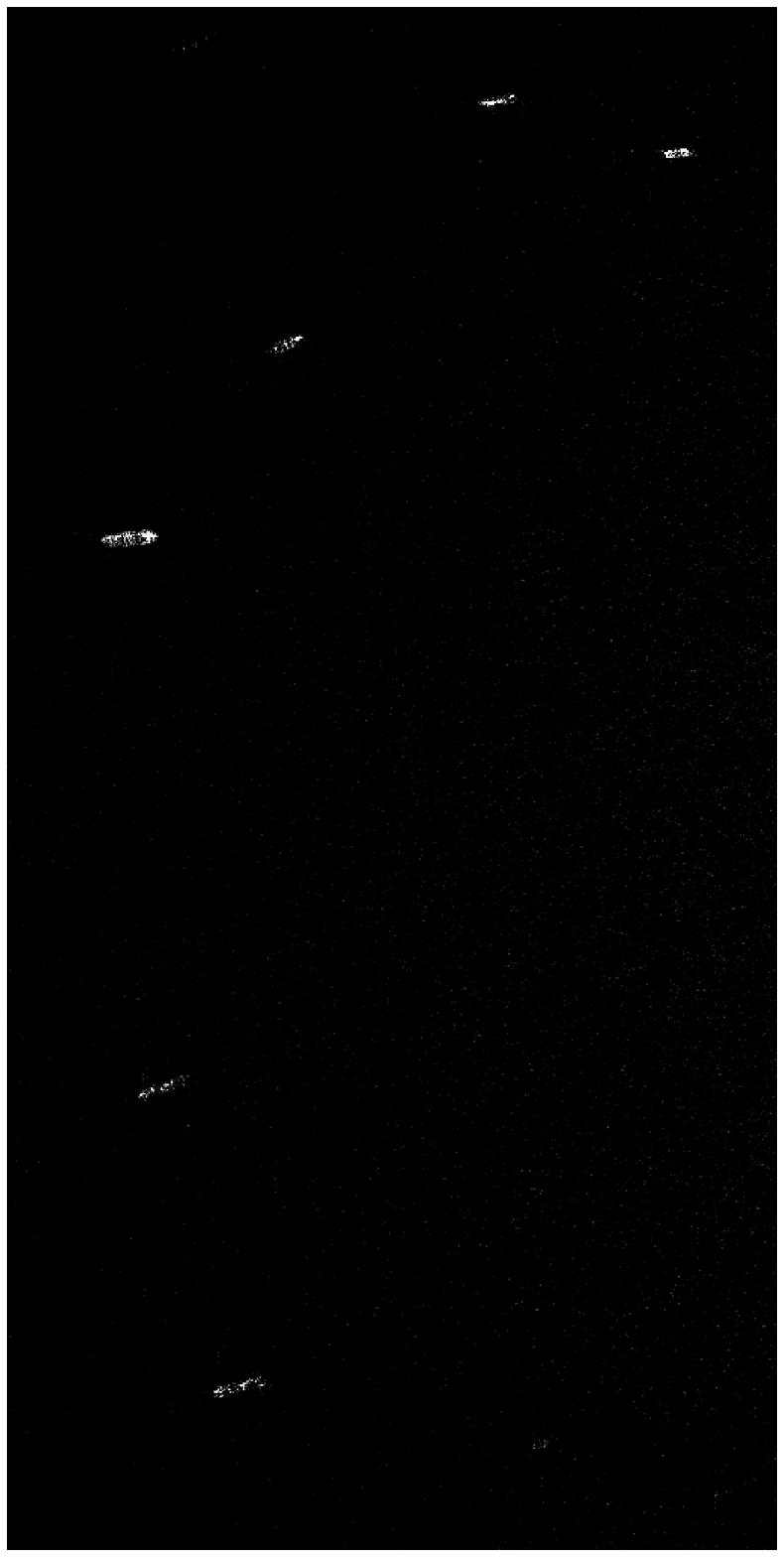}}
\hfil
\subfigure[CSEO(20\%)]{\includegraphics[width=0.22\textwidth]{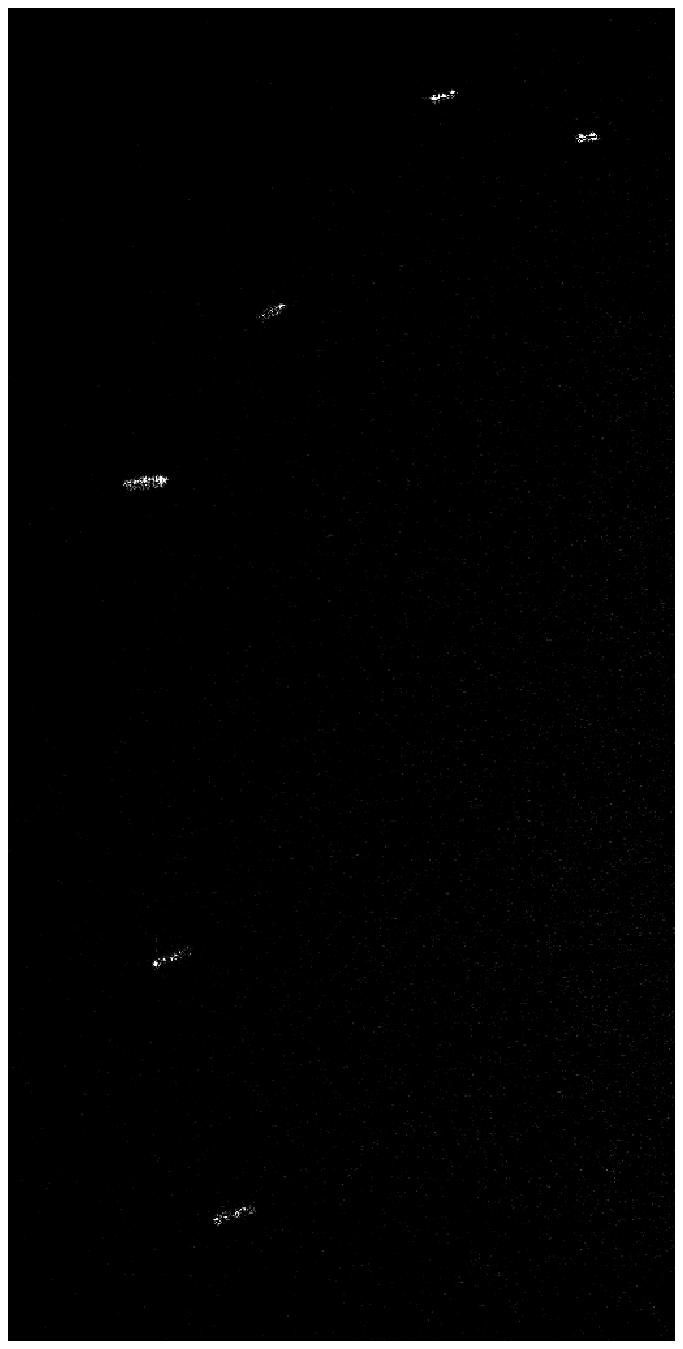}}
\hfil
\\
\centering
\subfigure[RDA(20\%)]{\includegraphics[width=0.22\textwidth]{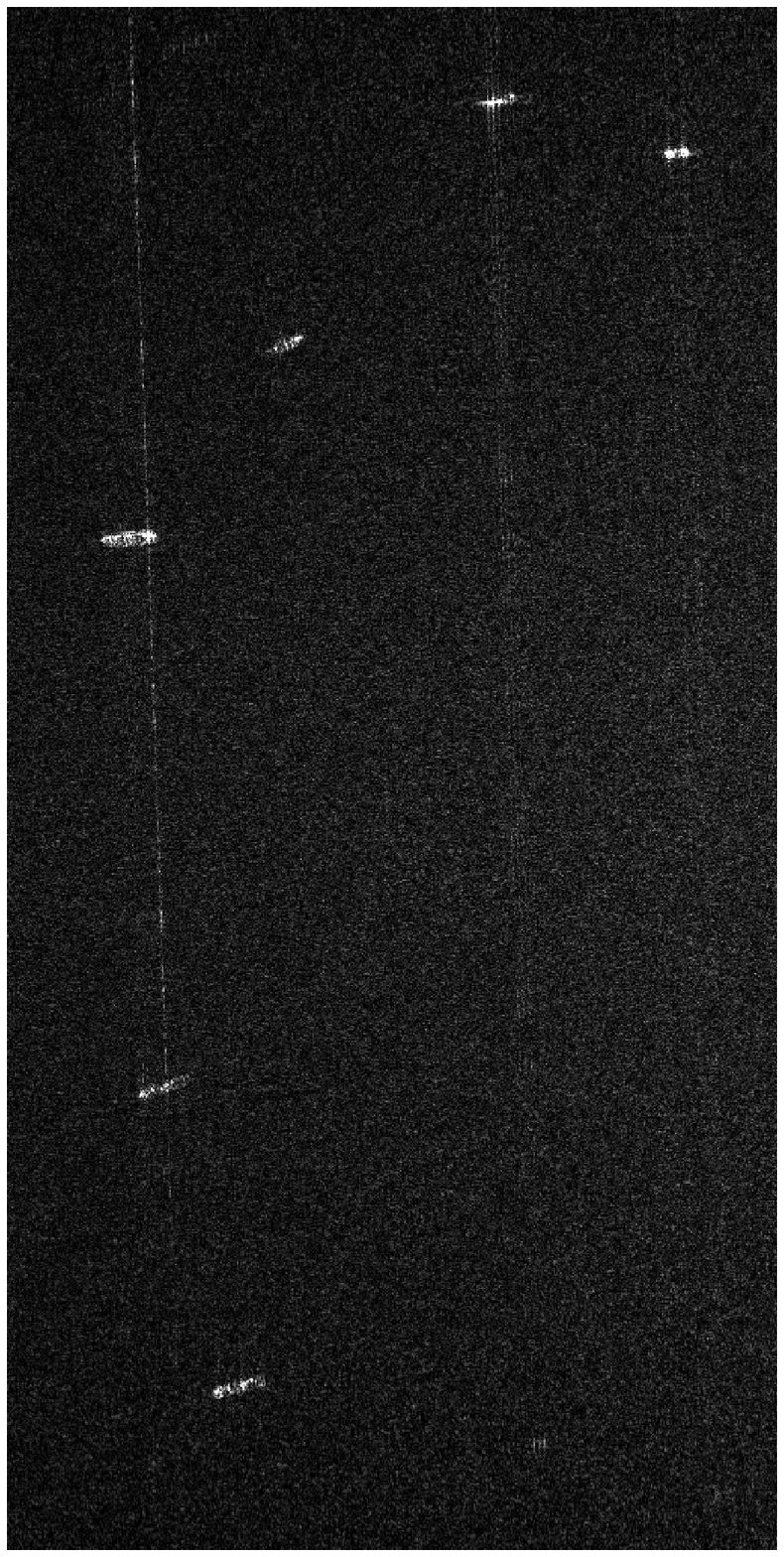}}
\hfil
\subfigure[CSRDA(5\%)]{\includegraphics[width=0.22\textwidth]{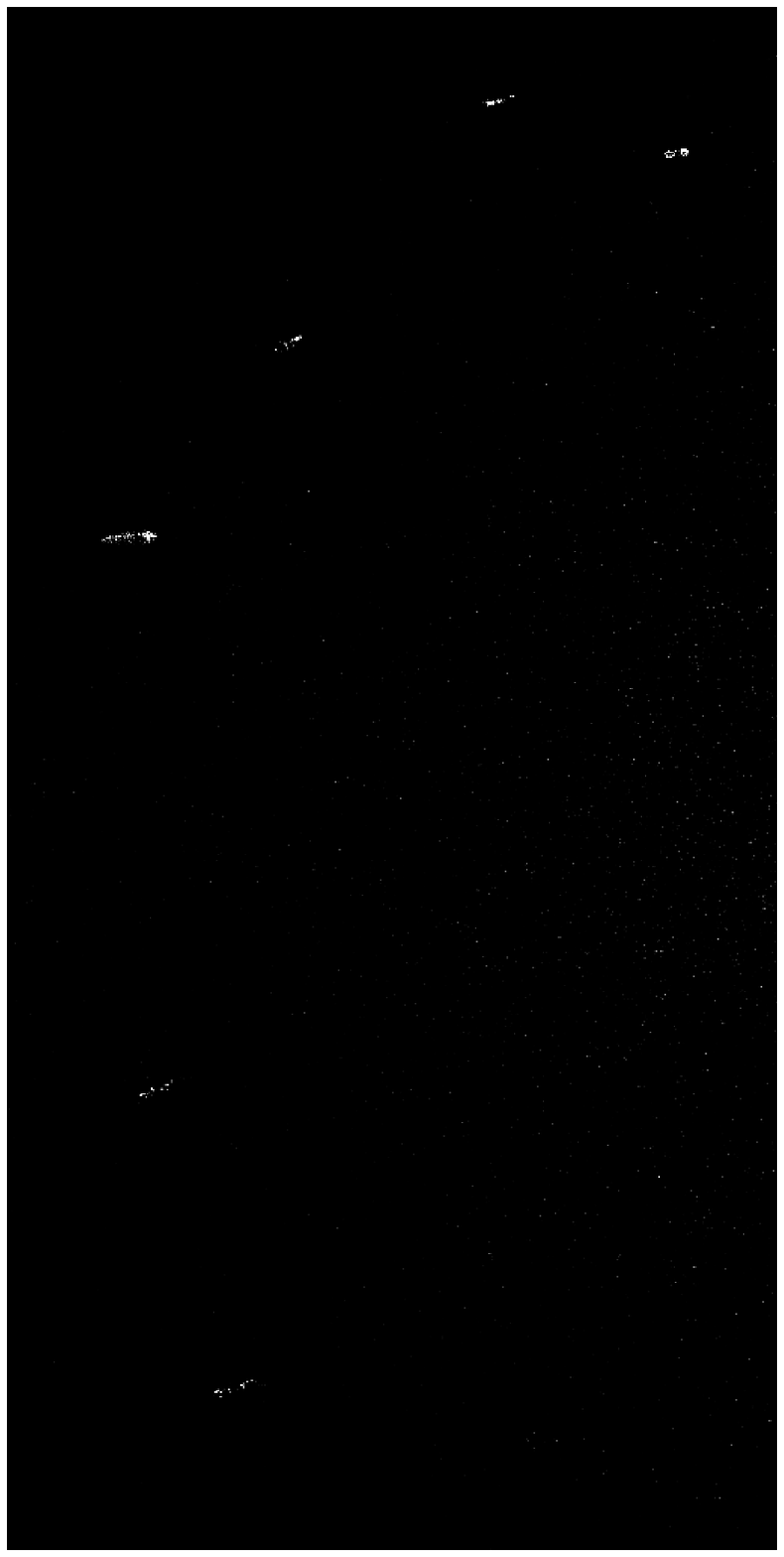}}
\hfil
\subfigure[CSEO(5\%)]{\includegraphics[width=0.22\textwidth]{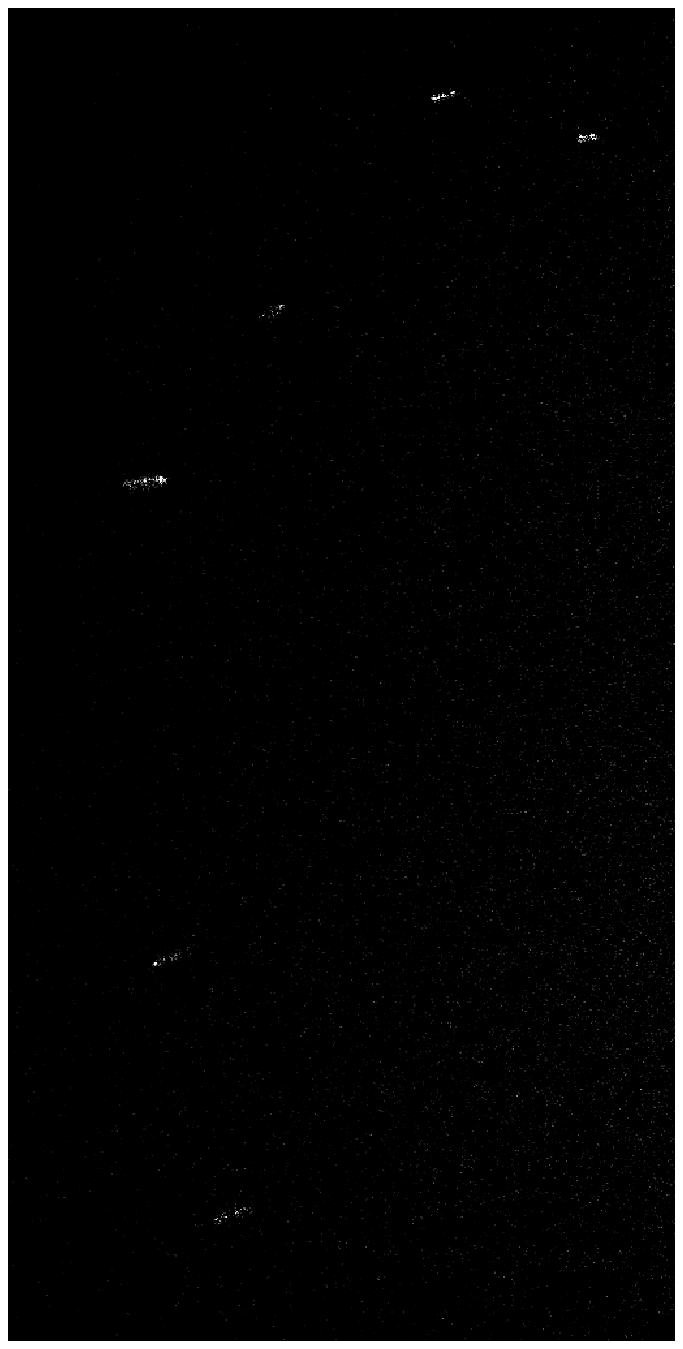}}
\hfil
\caption{Application on RADARSAT-1 (region of English Bay). (a) RDA with full samples. (b) CSEO with 20\% samples. (c) CSRDA with 20\% samples. (d) RDA with 20\% samples. (e) CSRDA with 5\% samples. (f) CSEO with 5\% samples. }
\label{fig_radar}
\end{figure}

All the above simulations support that the suggested approximated observation based CS-SAR method, CSRDA, is capable of high quality imaging under Nyquist rates and with a comparable complexity as the traditional MF based imaging method. Especially, as compared with the exact observation based CS-SAR imaging, CSRDA preserves the  features of imaging under Nyquist rate and reconstruction with feature enhancement, while reducing the imaging complexity dramatically. Such significant complexity reduction property makes the new method applicable to large scale imagery application (as will be demonstrated in the next subsection). This is, however, at the sacrifice of reconstruction quality or, equivalently, must be compensated with additional measurements. Thus, the new method provides a satisfying trade-off between the reconstruction complexity and quality.

\subsection{Application}
We have applied the new method, CSRDA, along with RDA and CSEO, to some real SAR imaging tasks. RADARSAT-1 is a famous satellite SAR launched at 1995, and the data used in this application were collected on June 16, 2002 with Fine Beam 2 about Vancouver region. The related key parameters of SAR system are as in Table. 1.

We first applied the 3 methods to reconstruction of the region of English Bay, in which 6 vessels are sparsely distributed, a very typical sparse scene. The scene was digitalized as $1024\times 512$ image with azimuth resolution 9 m and range resolution 6 m. We then reconstruct the image by the 3 imaging methods with sparsity $K=10000$, and with varied sampling rates from 100\% to 5\%. Some typical results of reconstructions are shown in Fig. \ref{fig_radar}. As expected, the application shows a completely similar performance as that in the simulations. For example, Fig. \ref{fig_radar}(a) exhibits that RDA can only reconstruct the image when samples are fully adopted, but strong side lobe is observed. When sampling rate is 20\%, RDA fails with obvious ambiguities, but CSRDA and CSEO both can perfectly recover the image, with much reduced side lobe. Fig. \ref{fig_radar}(c)(f) then show that when sampling rate is reduced to 5\%, the reconstruction of CSEO is with slightly higher precision than that of CSRDA, though both can still recover the targets. On the other hand, as listed in Fig. \ref{fig_RC}, CSRDA exhibits its dominant advantage in computational cost as compared with CSEO. For example, the computation time of reconstruction, when 20\% samples are used, by CSRDA is 1 minutes, while by CSEO is about 9 hours. Also, CSEO needs to store the sensing matrix which occupies about 16Gb memory, while only 100Mb for CSRDA.
\begin{figure}[t]
\centering
\subfigure[\ ]{\includegraphics[width=0.4\textwidth]{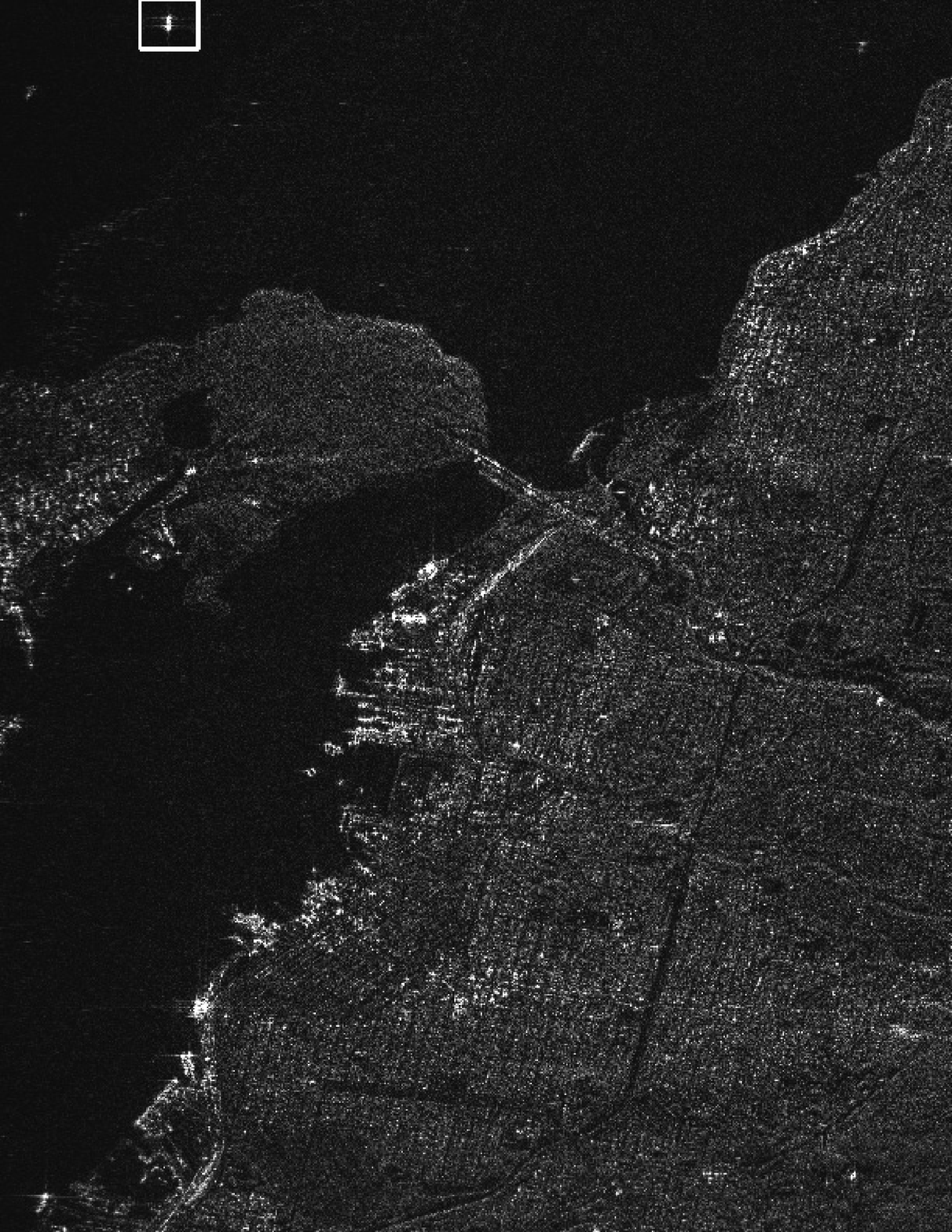}}
\hfil
\subfigure[\ ]{\includegraphics[width=0.4\textwidth]{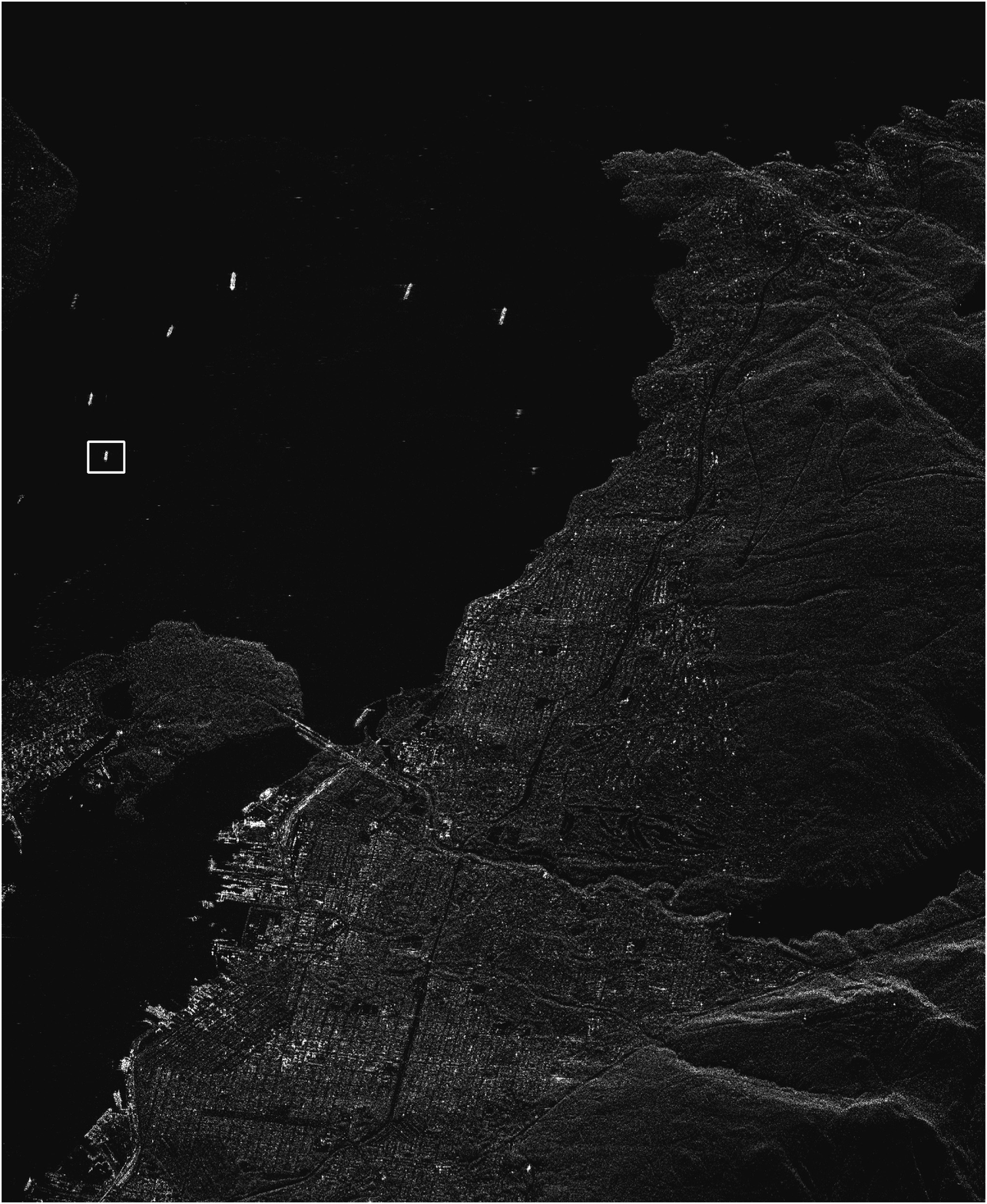}}
\hfil
\caption{Application results on RADARSAT-1. (a) RDA with full samples. (b) CSRDA with full samples. }
\label{fig_radarl}
\end{figure}
\begin{figure}[t]
\centering
\subfigure[\ ]{\includegraphics[width=0.3\textwidth]{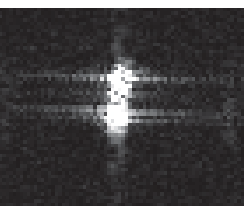}}
\hfil
\subfigure[\ ]{\includegraphics[width=0.3\textwidth]{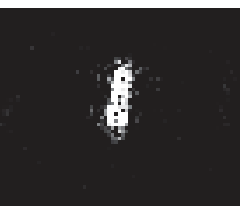}}
\hfil
\caption{Detailed comparison on the selected area with enlarged scale. (a) RDA. (b) CSRDA. }
\label{fig_radar2}
\end{figure}
We further applied the CSRDA to the large scale imaging problem together with RDA. However, the CSEO is not compared in this experiment because the memory cost is beyond the computational ability of our computer. The scene has a size $2048\times 2500$ samples, which is of large scale but not so sparse. CSRDA can be applied in principle because sparse regularization, which is adopted in CS-SAR, can be used as a feature enhanced imaging method (with suppressed side lobe and improved resolution), as demonstrated in the simulations. The reconstruction results by RDA and CSRDA (with 100\% sampling rate) are shown in Fig. \ref{fig_radarl}. It is seen from Fig. \ref{fig_radarl} that CSRDA has resulted the reconstruction with improved resolution and reduced side lobes, as demonstrated in the zoomed Fig. \ref{fig_radar2}.

The applications above support that the suggested approximated observation based CS-SAR imaging is effective and efficient, especially applicable to high dimensional SAR imaging applications. Such benefit clearly improves on the currently used exact observation based CS-SAR imaging methods.

\section{Conclusion}

Compressed sensing (CS) has been applied to yield novel SAR imaging methodologies under Nyquist sampling in recent years. The resultant CS-SAR models are time domain based and using the exact observation, which then makes it of very high computational cost, and is difficult to be applied in high dimensional applications. In this paper, we have proposed an approximated observation and frequency domain based CS-SAR imaging method, with which the computational complexity can be dramatically reduced.

The main contributions of the present work are as follows.

i) Instead of the exact observation matrix, an operator, called the approximated observation is constructed to generate SAR raw data by means of inverse of any traditional MF based imaging procedure (like RDA). Such generic construction makes the SAR imaging approximately decoupled, very fast in processing, while naturally connecting the existing MF based SAR imaging algorithms.

ii) Incorporating the approximated observation into CS-SAR framework, an efficient sparse regularization based CS-SAR model is formulated. The new model combines naturally CS and MF and is compatible with most existing SAR imaging methods, which needs a little modification of the current SAR imaging technologies.

iii) With the use of approximated observation, an iterative thresholding algorithm is suggested for fast solution of the new CS-SAR model, which forms a low complexity, CS featured, new SAR imaging method.

We have tested and applied the new suggested CS-SAR method with a series of simulations and applications. The experiments consistently support that the new method is capable of reconstructing sparse scenes with far fewer measurements than Nyquist requires, yielding always a feature enhanced  high quality imaging and bringing a speed up of reconstruction hundreds times as compared with the exact observation based CS-SAR methods. Due to the fast and feature enhancement features, the new CS-SAR method can be accepted as an efficient CS type SAR imaging technique, especially for high dimensional imaging applications.

It is worthwhile, however, to remark that although significant complexity reduction (speed-up of reconstruction) can be brought, the use of the approximated observation requires more samples to reconstruct a scene. Thus, how and what extent does the approximation affect the reconstruction deserves a further study. Moreover, since there are many possibilities of concrete realizations of the approximated observation, the criterion on how to select an appropriate one deserved study. All those problems are under our current research.

\appendices


\section*{Acknowledgment}
This work was supported by the State Key Development Program for Basic Research of China (973 Program)
(Grant No.2010CB731905),
Key Program of National Natural Science Foundation of China
(Grant No. 11131006),
and the National Natural Science Foundations of China
(Grants No. 61075054, 60975036, 11171272)

\ifCLASSOPTIONcaptionsoff
  \newpage
\fi



%

\bibliographystyle{./IEEEtran}
\bibliography{./IEEEabrv,./Mybib}

\begin{thebibliography}{10}
\providecommand{\url}[1]{#1}
\csname url@samestyle\endcsname
\providecommand{\newblock}{\relax}
\providecommand{\bibinfo}[2]{#2}
\providecommand{\BIBentrySTDinterwordspacing}{\spaceskip=0pt\relax}
\providecommand{\BIBentryALTinterwordstretchfactor}{4}
\providecommand{\BIBentryALTinterwordspacing}{\spaceskip=\fontdimen2\font plus
\BIBentryALTinterwordstretchfactor\fontdimen3\font minus
  \fontdimen4\font\relax}
\providecommand{\BIBforeignlanguage}[2]{{%
\expandafter\ifx\csname l@#1\endcsname\relax
\typeout{** WARNING: IEEEtran.bst: No hyphenation pattern has been}%
\typeout{** loaded for the language `#1'. Using the pattern for}%
\typeout{** the default language instead.}%
\else
\language=\csname l@#1\endcsname
\fi
#2}}
\providecommand{\BIBdecl}{\relax}
\BIBdecl

\bibitem{Cumming2004}
I.~G. Cumming, F.~H. Wong, U.~of~British~Columbia, and M.~D. Dettwiler,
  \emph{Digital Signal Processing of Synthetic Aperture Radar Data: Algorithms
  and Implementation}.\hskip 1em plus 0.5em minus 0.4em\relax Artech House,
  2004.

\bibitem{Candes2006cs}
E.~J. Cand{\`e}s, ``Compressive sampling,'' in \emph{Proceedings on the
  International Congress of Mathematicians}, 2006, pp. 1433--1452.

\bibitem{Baraniuk2007}
R.~G. Baraniuk, ``Compressive sensing,'' \emph{IEEE Signal Processing
  Magazine,}, vol.~24, no.~4, pp. 118--121, 2007.

\bibitem{Donoho2006}
D.~L. Donoho, ``Compressed sensing,'' \emph{IEEE Transactions on Information
  Theory}, vol.~52, no.~4, pp. 1289--1306, 2006.

\bibitem{Gurbuz2009}
A.~C. Gurbuz, J.~H. McClellan, and W.~R. Scott~Jr, ``Compressive sensing for
  subsurface imaging using ground penetrating radar,'' \emph{Signal
  Processing}, vol.~89, no.~10, pp. 1959--1972, 2009.

\bibitem{Herman2009}
M.~A. Herman and T.~Strohmer, ``High-resolution radar via compressed sensing,''
  \emph{IEEE Transactions on Signal Processing}, vol.~57, no.~6, pp.
  2275--2284, 2009.

\bibitem{Ender2010}
J.~H.~G. Ender, ``On compressive sensing applied to radar,'' \emph{Signal
  Processing}, vol.~90, no.~5, pp. 1402--1414, 2010.

\bibitem{Potter2010}
L.~C. Potter, E.~Ertin, J.~T. Parker, and M.~Cetin, ``Sparsity and compressed
  sensing in radar imaging,'' \emph{Proceedings of the IEEE}, vol.~98, no.~6,
  pp. 1006--1020, 2010.

\bibitem{Bhattacharya2007}
S.~Bhattacharya, T.~Blumensath, B.~Mulgrew, and M.~Davies, ``Fast encoding of
  synthetic aperture radar raw data using compressed sensing,'' in
  \emph{IEEE/SP 14th Workshop on Statistical Signal Processing}, 2007, pp.
  448--452.

\bibitem{Rilling2009}
G.~Rilling, M.~Davies, B.~Mulgrew \emph{et~al.}, ``Compressed sensing based
  compression of \protect{SAR} raw data,'' 2009.

\bibitem{Tello2010}
M.~Tello~Alonso, P.~L{\'o}pez-Dekker, and J.~J. Mallorqui, ``A novel strategy
  for radar imaging based on compressive sensing,'' \emph{IEEE Transactions on
  Geoscience and Remote Sensing}, vol.~48, no.~12, pp. 4285--4295, 2010.

\bibitem{Patel2010}
V.~M. Patel, G.~R. Easley, D.~M. Healy, and R.~Chellappa, ``Compressed
  synthetic aperture radar,'' \emph{IEEE Journal of Selected Topics in Signal
  Processing}, vol.~4, no.~2, pp. 244--254, 2010.

\bibitem{Zeng2011}
J.~S. Zeng, J.~Fang, and Z.~B. Xu, ``Sparse \protect{SAR} based on $l_{1/2}$
  regularization,'' in \emph{SCIENCE CHINA Information Sciences}, 2012(In
  press).

\bibitem{Cetin2001}
M.~{\c{C}}etin and W.~C. Karl, ``Feature-enhanced synthetic aperture radar
  image formation based on nonquadratic regularization,'' \emph{IEEE
  Transactions on Image Processing}, vol.~10, no.~4, pp. 623--631, 2001.

\bibitem{Khwaja2005}
A.~S. Khwaja, L.~Ferro-Famil, and E.~Pottier, ``\protect{SAR} raw data
  simulation using high precision focusing methods,'' in \emph{EURDA}, 2005,
  pp. 33--36.

\bibitem{Franceschetti2004}
G.~Franceschetti, R.~Guida, A.~Iodice, D.~Riccio, and G.~Ruello, ``Efficient
  simulation of hybrid stripmap/spotlight \protect{SAR} raw signals from
  extended scenes,'' \emph{IEEE Transactions on Geoscience and Remote Sensing},
  vol.~42, no.~11, pp. 2385--2396, 2004.

\bibitem{Wu1982}
C.~Wu, K.~Y. Liu, and M.~Jin, ``Modeling and a correlation algorithm for
  spaceborne \protect{SAR} signals,'' \emph{IEEE Transactions on Aerospace and
  Electronic Systems}, no.~5, pp. 563--575, 1982.

\bibitem{Wei2010}
S.~J. Wei, X.~L. Zhang, J.~Shi, and G.~Xiang, ``Sparse reconstruction for
  \protect{SAR} imaging based on compressed sensing,'' \emph{Progress In
  Electromagnetics Research}, vol. 109, pp. 63--81, 2010.

\bibitem{Candes2006}
E.~J. Cand{\`e}s, J.~Romberg, and T.~Tao, ``Robust uncertainty principles:
  Exact signal reconstruction from highly incomplete frequency information,''
  \emph{IEEE Transactions on Information Theory}, vol.~52, no.~2, pp. 489--509,
  2006.

\bibitem{Daubechies2004}
I.~Daubechies, M.~Defrise, and C.~De~Mol, ``An iterative thresholding algorithm
  for linear inverse problems with a sparsity constraint,''
  \emph{Communications on pure and applied mathematics}, vol.~57, no.~11, pp.
  1413--1457, 2004.

\bibitem{Xu2010}
Z.~B. Xu, X.~Y. Chang, F.~M. Xu, and H.~Zhang, ``\protect{$L_{1/2}$}
  regularization: a thresholding representation theory and a fast solver,''
  \emph{IEEE Transaction on Neural Networks and Learning Systems}, 2012(In
  press).

\bibitem{Blumensath2009}
T.~Blumensath and M.~E. Davies, ``Iterative hard thresholding for compressed
  sensing,'' \emph{Applied and Computational Harmonic Analysis}, vol.~27,
  no.~3, pp. 265--274, 2009.

\bibitem{Wu1976}
C.~Wu, ``A digital system to produce imagery from \protect{SAR} data,'' in
  \emph{Systems Design Driven by Sensors}, vol.~1, 1976.

\bibitem{Blumensath2010}
T.~Blumensath and M.~E. Davies, ``Normalized iterative hard thresholding:
  Guaranteed stability and performance,'' \emph{IEEE Journal of Selected Topics
  in Signal Processing}, vol.~4, no.~2, pp. 298--309, 2010.

\bibitem{Tropp2010}
J.~A. Tropp, J.~N. Laska, M.~F. Duarte, J.~K. Romberg, and R.~G. Baraniuk,
  ``Beyond \protect{N}yquist: Efficient sampling of sparse bandlimited
  signals,'' \emph{IEEE Transactions on Information Theory}, vol.~56, no.~1,
  pp. 520--544, 2010.

\end{thebibliography}

%


%
%




\end{document}